\documentclass{aa}  

\usepackage{graphicx}
\usepackage{txfonts}
\usepackage{hyperref}
\usepackage{breakurl}
\usepackage{natbib}
 \bibpunct{(}{)}{;}{a}{}{,} % to follow the A&A style
 \usepackage{amssymb}
\usepackage[export]{adjustbox}

\def\apjl{ApJL }
\def\aj{AJ }
\def\apj{ApJ }
\def\pasp{PASP }

\def\apjs{ApJS }
\def\araa{ARAA }
\def\aap{A\&A }

\def\mnras{MNRAS }

\def\memsai{Mem.~Soc.~Astron.~Italiana}

\def\cent{\mathrm{cm}}

\begin{document}

   \title{Absorption and scattering by interstellar dust in the silicon K-edge of GX~5-1}
   \titlerunning{The silicon K-edge of GX~5-1}

   \subtitle{}

   \author{S.T. Zeegers\inst{1,2},
          \,
          E. Costantini\inst{1},
          \,
          C.P. de Vries\inst{1},
          \,
          A.G.G.M. Tielens\inst{2},
          \, 
          H. Chihara\inst{3},
          \,
          F. de Groot\inst{4},
          \,         
          H. Mutschke\inst{5},
          \,
          L.B.F.M. Waters\inst{1,6}
          \,\&
          S. Zeidler\inst{7}
          }
    \authorrunning{S.T. Zeegers, E. Costantini, C.P. de Vries, A.G.G.M. Tielens}

   \institute{SRON Netherlands Institute for Space Research,
              Sorbonnelaan 2, 3584 CA Utrecht\\
              \email{S.T.Zeegers@sron.nl}
         \and
	      Leiden Observatory, Leiden University, PO Box 9513, NL-2300 RA Leiden, the Netherlands
         \and
         Department of Earth and Space Science, Osaka University, 1-1 Machikaneyama, Toyonaka, Osaka 560-0043 
         \and
         Debye Institute for Nanomaterials Science, 
         Utrecht University, Universiteitsweg 99, 3584 CG Utrecht, Netherlands   
         \and
         Astrophysikalisches Institut und Universit\"{a}ts-Sternwarte (AIU), 
         Schillerg\"{a}\ss{}chen 2-3, 07745 Jena, Germany
         \and
         Anton Pannekoek Astronomical Institute, 
         University of Amsterdam, P.O. Box 94249, 1090 GE Amsterdam, The Netherlands
         \and
         National Astronomical Observatory of Japan (NAOJ), 2-21-2 Osawa, Mitaka, Tokyo 181-8588, Japan
  }

   \date{Received 14 March 2016 / Accepted 14 December 2016}

% \abstract{}{}{}{}{} 
% 5 {} token are mandatory
 
  \abstract
   {We study the absorption and scattering of X-ray radiation by interstellar dust particles, which 
   allows us to access the physical and chemical properties of dust. The interstellar dust composition is not well understood, 
   especially on the densest sight lines of the Galactic Plane. X-rays provide a powerful tool in this study. }
   {We present newly acquired laboratory measurements of silicate compounds taken at the Soleil synchrotron 
   facility in Paris using the Lucia beamline. The dust absorption profiles resulting from this campaign were used 
   in this pilot study to model the absorption by interstellar dust along the line of sight 
   of the low-mass X-ray binary (LMXB) GX~5-1.
   }
   {The measured laboratory cross-sections were adapted for astrophysical data analysis and the
   resulting extinction profiles of the Si K-edge were implemented in the SPEX spectral fitting program. We derive the 
   properties of the interstellar dust along the line of sight by fitting the Si K-edge seen in absorption in the spectrum of GX~5-1.}
   {We measured the hydrogen column density towards GX~5-1 to be $3.40\pm0.1\times10^{22}\,\mathrm{cm}^{-2}$.
   The best fit of the silicon edge in the spectrum of GX~5-1 is obtained by a mixture of olivine and pyroxene.
   In this study, our modeling is limited to Si absorption by silicates with different Mg:Fe ratios.
   We obtained an abundance of silicon in dust of $4.0\pm0.3\times10^{-5}$ per H atom and a lower limit for
   total abundance, considering both gas and dust, of $>4.4\times10^{-5}$ per H atom,
   which leads to a gas to dust ratio of $>0.22$. 
   Furthermore, an enhanced scattering feature in the Si K-edge may suggest the presence 
   of large particles along the line of sight.}
   {}  

   \keywords{astrochemistry -- 
   X-rays: binaries -- dust, 
   extinction -- X-rays: individuals: GX~5-1
               }

   \maketitle
%
%________________________________________________________________

\section{Introduction}

Cosmic silicates form an important component of the dust present in the interstellar medium (ISM). 
These silicate dust particles are thought to be mainly produced in oxygen-rich asymptotic giant branch (AGB) stars \citep[e.g.,][]{Gail09}. 
Besides AGB stars, other sources such as novae, supernovae type II \citep{Wooden93,Rho08,Rho09}, 
young stellar objects \citep{Dwek80}, and red giant stars \citep{Nittler1997}
can produce silicate dust. Even dust formation in the ISM may occur in interstellar clouds \citep{Jones11}. 
Although the amounts of dust contributed by these sources is still debated \citep{Meikle07,Jones11}, silicate dust is abundant in the ISM
and can be found in many different stages of the life cycle of stars \citep{Henning10}. 
The physical and chemical composition 
of silicate dust has traditionally been studied at various wavelengths ranging from the radio to the UV and at different sight lines 
across the Galaxy \citep{Draine01,Dwek04}. 
However, there are still many open questions about, for instance, the chemical composition of silicates
\citep{Li01,Gail10}, the production and destruction rate of dust \citep{Jones94,Jones96},
the amount of crystalline dust in the interstellar medium, \citep{Kemper04} and the particle size distribution and the shape
of dust grains \citep[e.g.,][]{Min06,Min081,Voshchinnikov06,Mutschke09}. 
Furthermore, it is not precisely known how the dust composition and the dust particle size distribution change in different 
regions throughout the Galaxy \citep{Chiar06,Min07}. 

Elements such as C, O, Fe, Si, and Mg appear to be under-abundant in the cold phase of the ISM
\citep{Jenkins09,Savage96}. The abundances of these elements relative to hydrogen were found to be 
less than in the Sun, the Solar system, or in nearby stars \citep{Draine03}.
The atoms, that appear to be missing, are thought to be locked up in dust.  
This is referred to as depletion from the gas phase, which is defined here
as the ratio of the dust abundance to the total amount of a given element.
A large fraction of C, O, Fe, Si, and Mg is therefore thought to be depleted
and locked up in dust \citep{Henning10,Savage96,Jenkins09}.
Aside from carbon, which is mostly present in dust in graphite and polycyclic
aromatic carbon \citep[e.g.,][]{Zubko04,Draine07,Tielens08},
these elements form the main constituents of cosmic silicates \citep{Mathis98}.
Silicon in dust is mainly present in the ISM in the form of silicates, although it may also exist, 
in relatively small percentages, in the form of SiC:
0.1\%, \citep{Kemper04}, 9-12\% \citep{Min07}. 
Mg and Fe oxides are observed in stellar spectra \citep{Posch02,Henning95}, but
there is no observational evidence for them in the 
diffuse ISM \citep{Whittet97,Chiar06}. 
However, these compounds have been isolated as stardust in Solar system meteorites \citep{Anders93}.

An important property of interstellar dust is crystallinity.  
From observations of the $10\,\mu\mathrm{m}$ and $18\,\mu\mathrm{m}$ features, \citet{Kemper04} 
concluded that along sight lines towards the Galactic center only $1.1\%$
(with a firm upper limit of $2.2\%$)
of the total amount of silicate dust has a crystalline structure. 
On the other hand, some of the interstellar dust grains captured by the Stardust Interstellar Dust Collector \citep{Westphal14}
showed a large fraction of crystalline material. The cores of these particles contained crystalline forsteritic olivine. 
\citet{Westphal14} conclude that crystalline materials are probably preserved in the interiors of larger ($>1\,\mu$m) particles.
Interestingly, dust is found to be in crystalline form at the start and at the end of the life cycle of stars, 
whereas very little crystalline dust appears to survive the harsh environment of the ISM. 
There are indications that the amount of crystalline dust depends on the environment. 
For instance, silicate dust in starburst galaxies appears in large fractions in crystalline form \citep{Spoon2006,Kemper2011}, 
probably reflecting freshly produced dust.
Indeed, silicates in the ISM can be amorphous either because during the formation process the silicates condense as amorphous grains 
\citep{Kemper04,Jones12}
or the crystal structure is destroyed in the ISM by cosmic ray bombardments, UV/X-ray radiation, and supernova shock waves 
\citep{Bringa07}. In the first case the silicates will have a non-stoichiometric composition and in the second case they will have
the stoichiometry of the former crystal \citep{Kemper04}. 
The dust features of amorphous dust are smoother than those of crystalline dust, 
which makes the determination of the structure and composition of the interstellar dust from spectral studies more difficult.

From X-ray observations of sight lines towards the Galactic plane and infrared observations towards the Galactic center,
silicates were found to be Mg-rich rather than Fe-rich 
\citep{Costantini05, Costantini2012, Lee09,Min07}.
However, Fe is heavily depleted ($70-99\%$) and probably mostly locked up in dust grains \citep{Wilms00,Whittet03}.
It is not certain in which exact form Fe is incorporated into dust 
\citep{Whittet97,Chiar06}.
Since the composition of certain silicates allows iron rich compounds, it is possible that some of the 
iron is locked up in these silicate grains.
Another and possibly complementary scenario to preserve Fe in dust 
prescribes that Fe could be locked up in Glass with Embedded Metal and Sulfides (GEMS)
of interstellar origin \citep[e.g.][]{Bradley94, Floss06, Keller13}. 

The abundances of most of the important metals decrease with distance from the Galactic Plane, 
which can be described by a gradient with an average slope of 
$~0.06$ dex $\mathrm{kpc}^{-1}$ \citep[and references therein]{Chen03}.
Although the ISM shows this general gradient, the ISM is also very patchy. 
The measurements of abundances show a large scatter as function of the Galactic radius, due to local influences of, for
instance, supernova ejecta and 
infalling metal-poor gas onto the disk \citep{Nittler05}. % misschien uit Nittler alles citeren...
The $10\mu\mathrm{m}$ feature provides information 
about the Si abundance in the Galaxy, which in turn can provide restrictions on the dust composition
and possibly on the dust size distribution \citep{Tielens96}.
It is not precisely known how the dust distribution and the 
dust composition change relative to the environment. 
Simple dust size distributions, 
such as the Mathis-Rumpl-Nordsieck (MRN, \citet{MRN77}) distribution, consisting of solid spherical dust particles, 
may not be sufficient to explain the observations towards dense regions of the Galaxy. 
Dust particles may be non-spherical and porous due to the formation processes of dust \citep{Min06,Chiar06,Min07}.
Furthermore, \citet{Hoffman16} show the importance of incorporating non-spherical dust particles into X-ray scattering 
analyses. 

X-ray observations provide a hitherto relatively unexplored but powerful probe of interstellar 
dust \citep{Draine03, Lee09,Costantini2012}. The extinction features near the X-ray edges of 
O, Mg, Si, and Fe can be analyzed depending on 
the column density on the line of sight towards the source and the sensitivity of the detector. 
The X-ray Absorption Fine structures (XAFS) near the atomic absorption edges of elements provide a unique 
fingerprint of the dust. 
These XAFS have been observed in the X-ray spectra of astrophysical objects in data 
from \textit{XMM} and \textit{Chandra} \citep{Lee01,Ueda05,Kaastra09,Devries09,Pinto10,Pinto13,Costantini2012,Valencic13}. 
The X-rays provide important advantages compared to longer wavelengths and, in that way, provide an independent 
method to study silicate dust. The two most important advantages are that it is possible to measure the 
quantity of absorbing gas and dust simultaneously and to directly determine the composition of the dust. 
In particular, it is, in principle, possible to address the abundance, composition, 
stoichiometry, crystallinity, and size of interstellar silicates.

Bright X-ray binaries, distributed along the Galactic Plane, 
can be used as background sources to probe the intervening dust and gas in ISM along the line of sight. 
In this way, a large range of column densities can be investigated and it is possible to analyze dust in various regions in the Galaxy. 
Dust in diffuse regions along the Galactic plane has been studied in the X-rays by \citet{Lee09}, \citet{Pinto10}, 
\citet{Costantini2012}, and \citet{Pinto13} for several sight lines. 
The dense ISM has been less extensively studied in the X-rays. These dense environments 
(hydrogen column density $>1\times10^{22}\,\mathrm{cm}^{-2}$) can be studied in the X-rays by 
observing the Mg and Si K-edge. In this paper we will focus in the Si K-edge. 
 
The information on silicon in silicates of astronomical interest is very limited and sparse. The Si K-edge of some silicates
has been measured by, for example, \citet{Li94,Poe97}, but most of these silicates cannot be used in astronomical studies
because they also contain elements which are not abundant in the ISM.
In this first study of the Si K-edge in astronomical data with a physically motivated model, we present a new set of laboratory measurements of 
Si K-edges of six silicate dust samples. 
The samples contain both crystalline and amorphous silicates.
Further details about these samples are given in section 3.1. 
The measurements are part
of a large laboratory measurement campaign aimed at the characterization of interstellar dust analogs \citep{Costantini13}.

We analyze the interstellar matter along the line of sight of X-ray binary GX~5-1 
using models based on new laboratory measurements and discuss the properties and composition of the dust. 
The paper is structured in the following way: in section 2, we explain the usage of XAFS to study interstellar dust. 
In section 3, the data analysis of the laboratory samples is described. Section 4 shows the calculation of the extinction 
cross-section (absorption and scattering) of the samples. Section 5 describes the source GX~5-1.
In section 6, we fit the models to the spectrum of GX~5-1 and determine the best fit to the data 
using the samples from section 3. We discuss our results in section 7 and conclude in section 8.

%__________________________________________________________________

\section{X-ray absorption edges}

The modulations at the absorption edges of elements locked up in dust are called X-ray absorption fine structures
\citep[XAFS,][]{meurant1983solid}. XAFS are best understood in terms of the wave behavior of the photo-electron.
These structures arise when an X-ray photon excites a core electron. 
The outwardly propagating photo-electron wave will be scattered by the neighboring atoms. 
From these atoms new waves will emanate and 
will be superimposed on the wave function of the photo-electron.  
The wave function of the scattered photo-electron is therefore modified due to constructive and destructive interference. In this way, the 
absorption probability is modified in a unique manner, because it depends on the configuration of the neighboring atoms.
These modulations can be used to determine the structure of the silicate dust in the ISM, because the modulations show unique features for 
different types of dust.  
 
\section{Laboratory data analysis}

\subsection{The samples}

We analyzed six samples of silicates.
The compounds are presented in Table~\ref{table:elements}. Three of the samples were
natural crystals, that is, two orthopyroxenes, one of
them magnesium-rich (sample 4: enstatite, origin Kiloza, Tanzania),
one with a higher iron content (sample 6: hypersthene, origin:
Paul Island, Labrador), and one is an iron-rich olivine
(sample1: olivine, origin: Sri Lanka). 
See also \citet{Jaeger98}
and \citet{Olofsson2012} for infrared data of the
enstatite and the olivine crystals.

The three other samples, that is, sample 2, sample 3 (which is the crystalline counterpart of sample 2), and sample 5,
were synthesized for this analysis in
laboratories at AIU Jena and Osaka University.
The amorphous $\mathrm{Mg}_{0.9}\mathrm{Fe}_{0.1}\mathrm{SiO}_{3}$
sample has been synthesized by quenching of a melt
according to the procedure described by \citet{Dorschner95}. 
The crystalline counterpart was
obtained by slow cooling of silicate material
produced under Ar atmosphere in an electric arc,
similarly to that described for Mg/Fe oxides in
\citet{Henning95}.

We were motivated in the choice of the sample by the almost absolute absence of XAFS measurements for silicates of astronomical interest.
In order to produce laboratory analogs of interstellar dust silicates, four main criteria were considered: 

\begin{itemize}
\item{The samples (or a mixture of these samples) should reflect the interstellar dust silicates of "mean" cosmic composition.}
\item{The samples have an olivine or pyroxene stoichiometry.}
\item{The samples contain differences in the Mg:Fe ratio.}
\item{The sample set contains both amorphous and crystalline silicates.} 
\end{itemize}

The composition of the samples present in our study is chosen in such a way 
that mixtures of these samples can reflect the cosmic silicate mixture as described by \citet{Draine84}.
According to observations of 10 and 20$\,\mu$m feature in the infrared, the silicate dust mixture consists of 
an olivine and pyroxene stoichiometry \citep{Kemper04, Min07}. 
The dominating component seems to be silicates of an olivine stoichiometry \citep{Kemper04}. 
\citet{Min07} show that the stoichiometry lies in between that of olivine and pyroxene, 
which suggests a mixture of these two silicate types. Therefore, our sample set contains both pyroxenes and an olivine silicate. 
The samples show variations in the Mg:Fe ratio. 
These variations reflect the results from previous studies of interstellar dust. 
\citet{Kemper04} infer from the observed stellar extinction that $\mathrm{Mg}/(\mathrm{Mg}+\mathrm{Fe})\!\sim\!0.5$, 
whereas \citet{Min07} conclude that $\mathrm{Mg}/(\mathrm{Mg}+\mathrm{Fe})\!\sim\!0.9$. 
Silicates with a high magnesium fraction of $\mathrm{Mg}/(\mathrm{Mg}+\mathrm{Fe})\!\sim\!0.8$ have been found in 
environments that show silicates with a crystalline structure, for example 
around evolved stars \citep{Molster202,Molster102,Molster302,Devries10}, in comets, 
\citep{Wooden1999,Messenger05} and in disks around T Tauri stars \citep{Olofsson2009}. 
Therefore, our samples have ratios of Mg/(Mg+Fe) that vary between 0.5 and 0.9. 
The sample set also contains both amorphous and crystalline silicates.  

Besides the two amorphous pyroxene samples present in this set,
another highly suitable candidate would be an amorphous olivine, which is not present in this analysis.
In general, amorphous olivine is difficult to synthesize, because this particular silicate crystallizes extremely quickly.
A rapid-quenching technique is necessary to prevent the crystallization process.
For compounds with a higher iron content, this technique may not be fast enough to prevent both phase separation and crystallization.
Even at moderately high Fe contents, the silicates would show variations of the Mg:Fe ratio throughout 
the sample. These variations are problematic in the comparison with their crystalline counterpart. 
In the initial phase of this campaign, we indeed also analyzed the amorphous olivine, originally used in \citet{Dorschner95}. 
The inspection with the electron microscope revealed this particular sample to be both partially inhomogeneous
in composition and contaminated by tungsten. This element has significant M-edges around the Si K-edge (i.e., the M1-M5 edges of tungsten fall in a range
between 1809 and 2819 eV, where the Si K-edge is at 1839 eV). For these reasons, the amorphous olivine  
from \citet{Dorschner95} was discarded at a very early stage of this campaign, as it was unsuitable for fluorescence measurements in the X-rays.

\subsection{Analysis of laboratory data}

Ideally one would like to measure the absorption of the samples directly in transmission through the sample, because this would
resemble the situation in the ISM more closely. 
However, to measure the dust samples in transmission around the energy of the Si K-edge,
we need optically thin samples, which implies a sample thickness of 
1.0 - 0.5$\,\mu\mathrm{m}$. This is impossible to obtain for practical reasons. 
In our analysis of the silicon K-edge we make use of optically thick samples (i.e., the sample is much thicker than the penetration depth) 
and therefore it is necessary to use a different technique with which the absorption can be derived. 
There are two processes that can be used in this case, which occur after
the core electron is excited by an X-ray photon. The excited photo-electron 
leaves a core hole. This can be filled by an electron from a higher shell that falls into the vacancy. The excess energy 
can either be released as a fluorescent photon or another electron gets ejected. The latter is called the Auger effect \citep{Meitner1922}. 
Depending on the attenuation of the signal, both effects can be used to derive the amount of absorption around the edge. 
Since the fluorescent signal of our measurements was strong enough, 
we used the fluorescent measurements of the Si $\mathrm{K}_{\alpha}$ line in our analysis of the Si K-edge. 

The absorption (given here by the absorption coefficient $\alpha(E)$) 
can be derived from the fluorescent spectrum by dividing the fluorescent intensity ($I_f$) by the beam intensity ($I_0 $).
\begin{equation}
 \alpha(E)\propto I_f/I_0\
 \label{eq:fluor}
\end{equation}

The samples were analyzed at the Soleil synchrotron facility in Paris using the Lucia beamline. 
They were placed in the X-ray beam and the reflecting fluorescent
signal was measured by four silicon drift diode detectors. The beam has an energy range of 0.8 - 8 keV.
The energy source of the X-ray beam is an undulator, which creates a collimated beam. 
The beam is first focused by a spherical mirror and  
then passes two sets of planar mirrors that act as a lower pass filter. This procedure reduces the
high-order contamination from the undulator and the thermal load received by the monochromator crystals.
The monochromator crystals then rotate the beam and keep the exit beam at a constant height. There are five different crystals available.
During our measurements, we made use of the  KTP monochromatic crystals. 
In the energy range of 1280 to 2140 eV, the KTP monochromatic system has an energy resolution between 0.25 and 0.31 eV \citep{Flank2006}. 
The beam energy can be increased gradually (stepped by the resolution of the monochromator) 
in order to measure the absorption at the pre-edge (1800 - 1839 eV), the edge itself (at 1839 eV),
and the post-edge (1839 - 2400 eV). 
The measurements were made with 0.5 eV energy spacing between measurements close to the edge. 
The added signal of the four silicon drift diode detectors yields the total fluorescent spectrum from the sample, from which the 
absorption coefficient $\alpha(E)$ can be derived using the beam intensity $I_0$ as indicated in Equation~\ref{eq:fluor}. 

All six samples were stuck on two identical copper sample plates. The silicates were ground to a powder and pressed into 
a layer of indium foil, which made it 
possible for the samples to stick to the copper plates. Each sample was measured twice and 
therefore four measurements of each compound were obtained. This was done to avoid any dependence in the measurement
on the position of the sample on the copper plate. 
The average of the four measurements is used in this analysis. The dispersion among the measurements is small; $3\%$.

\begin{figure}
 \begin{center}
  \includegraphics[scale=0.52]{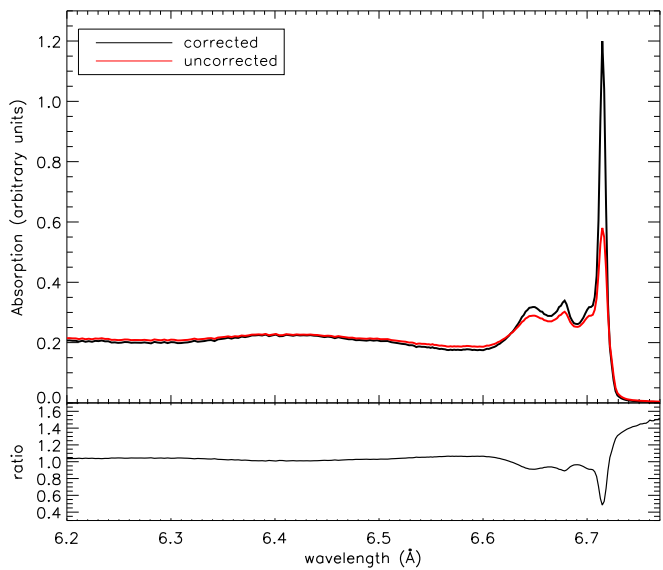}
 \caption{\small{Example of the difference between the sample corrected for saturation and the uncorrected measurement. Shown is sample 3,
 crystalline pyroxene. The lower panel shows the ratio of the corrected and uncorrected measurement.}}
  \label{fig:fluo_correction}
  \end{center}
\end{figure}

% Can be found in: /home/sascha/idl/idl_prog/plots_met_residuals.pro

The measurements of the samples were corrected for pile-up and saturation. Pile-up is caused by the detection of two photons instead of 
one at the same time on the detector. This can be seen on the detector as an extra fluorescent line at twice the energy of the expected fluorescent
$\mathrm{K}_\alpha$ line. 
This means that some of the intensity of the original line would be lost. To correct for this effect, we isolated both the silicon fluorescent
line and the associated pile-up line from the fluorescent spectrum. The contribution of the pile-up line is then weighed and added to the original line. 
A comparison between the corrected and uncorrected data shows that the influence of pile-up in the sample is minimal ($<1\%$).  

As indicated by Equation~\ref{eq:fluor}, the intensity of fluorescence is proportional to the absorption probability, but this is 
a slight oversimplification. The fluorescent light has to travel through the sample before it can be detected. On the way 
through the sample, the photons can be absorbed by the sample itself and the fluorescence intensity is attenuated. This effect is called 
saturation. This means that the measured fluorescent signal $I_f$ is no longer proportional to the absorption coefficient. 
In our measurements, the samples were all placed at an angle $\theta=45\,^{\circ}$; therefore, the angular dependence can be neglected. 
Another advantage of positioning the sample in this way is,
that the beam and the detector are now at a $90\,^{\circ}$ angle. Due to the polarisation of the radiation from the incident beam, 
the beam is greatly suppressed at this angle and almost no radiation from the beam can reach the detector directly. 
The measured fluorescence intensity divided by the intensity of the beam can now be expressed as: 

\begin{equation}
 \frac{I_f}{I_0}=\epsilon_{f}\frac{\Omega}{4\pi}\frac{\alpha_{e}(E)}{\alpha_{\mathrm{tot}}(E)+\alpha_{\mathrm{tot}}(E_f)}[1-e^{{-[\alpha_{\mathrm{tot}}(E)+\alpha_{\mathrm{tot}}(E_f)}]x_n/\sin(\theta)}]
\label{eq:flou1}
 \end{equation}

In this equation, $\epsilon_{f}$ is the fluorescence efficiency, $x_n/\sin(\theta)$ is the effective optical path
(where $x_n$ is the penetration depth into the sample and $\theta$ is the angle between the sample surface and the beam), 
$\Omega$ is the solid angle of the detector, $E_f$ is the energy 
of the fluorescent X-ray photons, $\alpha_{e}(E)$ is the absorption from the element of interest and $\alpha_{\mathrm{tot}}$ is the total
absorption defined as: $\alpha_{\mathrm{tot}}=\alpha_{e}(E)+\alpha_b(E)$. $\alpha_b(E)$ denotes the absorption from all other atoms and other
edges of interest. For concentrated samples, $\alpha_b(E)$ can become dominant and the XAFS will be damped by this saturation effect. 
In the ISM, the dust is 
very diluted, so saturation will not occur in real observations, but has to be corrected for in bulk matter.

In the case of a thick sample, the exponential term in Equation~\ref{eq:flou1} becomes small and can be ignored: 
\begin{equation}
 \frac{I_f}{I_0}=\epsilon_{f}\frac{\Omega}{4\pi}\frac{\alpha_{e}(E)}{\alpha_{\mathrm{tot}}(E)+\alpha_{\mathrm{tot}}(E_f)}
 \label{eq:flou2}
\end{equation}

The correction has been done using the FLUO software developed by Daniel Haskel. FLUO is part of the UWXAFS
software \citep{Stern1995117}.
A detailed explanation of this correction can be found in Appendix 1.  
\footnote{\burl{http://www.aps.anl.gov/~haskel/fluo.html}} % put in reference to Haskel
This routine uses tabulated absorption cross-sections to correct
most of the distortion in order to recover the actual absorption coefficient. 
In comparison to the correction for pile-up, the correction of the absorption spectra for saturation 
is considerable as can be seen in Figure~\ref{fig:fluo_correction}.

\begin{figure*}
 \begin{center}
 \includegraphics[scale=0.7]{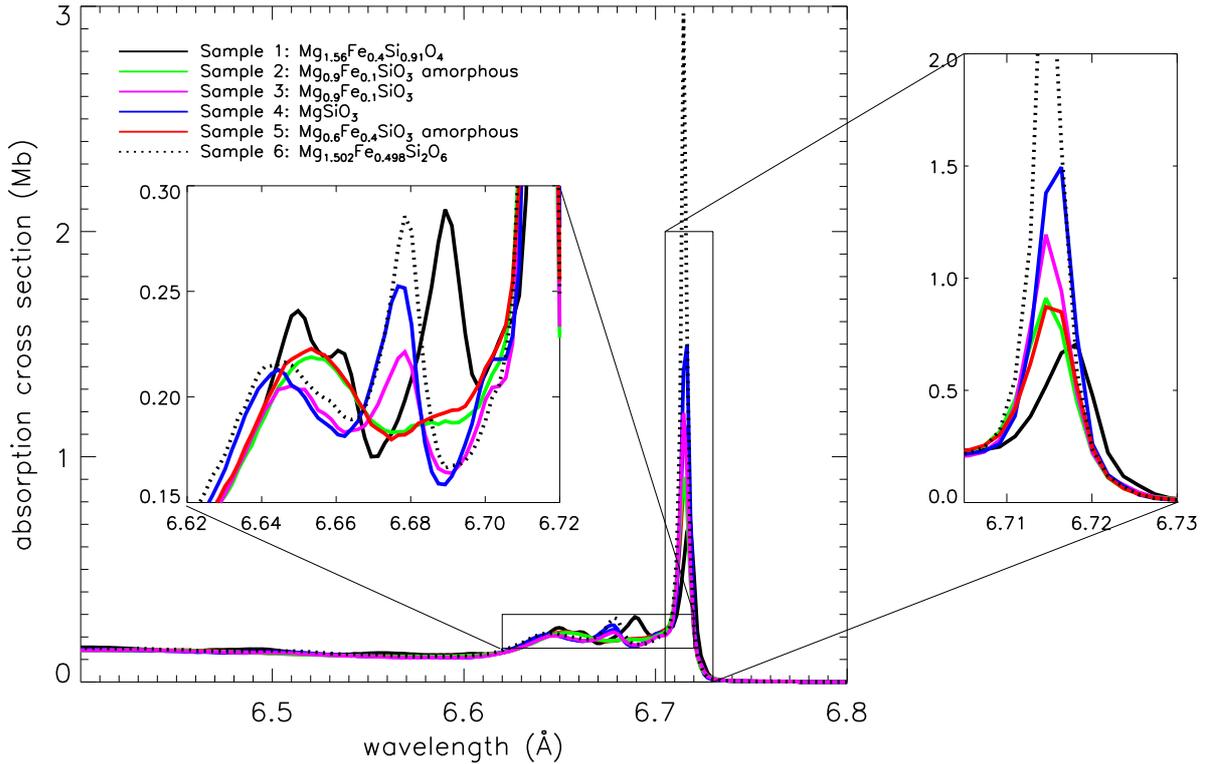}
 \caption{\small{The Si K-edge of the six samples. The x-axis shows the energy in $\AA$ and the 
 y-axis shows the amount of absorption indicated by the cross-section (in Mb per Si atom). }}
 \label{fig:crystal_and_amorphous}
  \end{center} 
\end{figure*} 
%Figure can be found in: ~/idl/idl_prog/plots_elisa_vidi_2015.pro

Figure~\ref{fig:crystal_and_amorphous} shows the amount of absorption, indicated by the cross-section (in arbitrary units),
corrected for saturation and pile-up
as a function of the energy of the silicon K-edge for all the samples. 
This figure shows the differences in the XANES for different compounds.  
The structure of olivine (sample 1), for example, shows a peak at $6.69\,\AA$ that is not observed in the pyroxene samples.  
There is also a clear difference between amorphous and crystalline samples. Indeed, the structures that are present between 
6.70 and 6.66$\,\AA$ in sample 3, for example, are washed out in the amorphous counterpart: sample 2. 
 
\begin{table}
\caption{Samples} % title of Table
\label{table:elements} % is used to refer this table in the text
\centering % used for centering table
\begin{tabular}{l l l l} % centered columns (4 columns)
\hline\hline % inserts double horizontal lines
No. sample & Name &Chemical formula& Structure  \\ % table heading
\hline \noalign{\smallskip}% inserts single horizontal line
1 & Olivine&$\mathrm{Mg}_{1.56}\mathrm{Fe}_{0.4}\mathrm{Si}_{0.91}\mathrm{O}_4$& crystal  \\ % inserting body of the table
2 & Pyroxene&$\mathrm{Mg}_{0.9}\mathrm{Fe}_{0.1}\mathrm{Si}\mathrm{O}_3$& amorphous  \\
3 & Pyroxene&$\mathrm{Mg}_{0.9}\mathrm{Fe}_{0.1}\mathrm{Si}\mathrm{O}_3$& crystal \\
4 & Enstatite&$\mathrm{Mg}\mathrm{Si}\mathrm{O}_3$& crystal$^*$   \\
5 & Pyroxene&$\mathrm{Mg}_{0.6}\mathrm{Fe}_{0.4}\mathrm{Si}\mathrm{O}_3$& amorphous\\
6 & Hypersthene&$\mathrm{Mg}_{1.502}\mathrm{Fe}_{0.498}\mathrm{Si}_{2}\mathrm{O}_6$& crystal\\
\hline %inserts single line
\end{tabular}
\tablefoot{$^*$Sample 4 contains a very small amount iron, which is not significant in our analysis. The Fe:Mg ratio is $4\times10^{-2}$. }
\end{table}

\section{Extinction cross-sections}

We derived the amount of absorption of each sample in arbitrary units from the laboratory data. 
These measurements need to be converted to extinction cross-sections (in units Mb) in order to
implement them into the AMOL model of the fitting routine SPEX \citep{Kaastra1996} for further analysis. 
The total extinction cross-section can be calculated by using the Mie theory~\citep{Mie1908}. 
In order to do so, we first need to derive the optical constants of the samples.  
In this section, we explain the methods used to 
obtain the extinction cross-section of each sample.  

\subsection{Optical constants}
When light travels through a material, it can be transmitted, absorbed, or scattered. The transmittance $T$ is defined as 
the ratio of transmitted $I$ and incident light $I_0$. The amount of light that is absorbed or transmitted depends on 
the distance the light travels through the material and on the properties of the material.
The transmittance is described by the Beer-Lambert law (Equation~\ref{eq:beer_lambert}):

\begin{equation}
 T=\frac{I}{I_0}=e^{-\alpha x} = e^{-x/l}
 \label{eq:beer_lambert}
\end{equation} 

In this equation, $x$ is the depth of the radiation in the material and $\alpha$ is the extinction coefficient. 
The transmittance is also equivalent to $e^{-x/l}$,
where $x$ is again the depth of the radiation into the material and $l$ is the mean free path (e.g., the average distance
travelled by a photon before it is absorbed). 
The extinction coefficient depends on the properties of the material and is independent of the distance $x$
traveled through medium. However, $\alpha$ does depend on the wavelength of the incident light. 
It can be expressed as $\alpha=\rho \kappa_{\lambda}$, where $\rho$ is the specific density of the material and $\kappa_{\lambda}$ is the 
cross-section per unit mass. 
The Beer Lambert law is an approximation, assuming that the reflections at the surfaces of the material are negligible. 
In the X-rays, the contribution of reflection becomes very small.  

In order to determine $\alpha$ from our measurements, we transform the absorption in arbitrary units obtained from the laboratory 
fluorescent measurements in section 3, to a transmission spectrum. In order to do this, we use the tabulated values of the mean free path
$l$ provided by 
the Center for X-ray Optics at Lawrence Berkeley National laboratory\footnote{\burl{http://www.cxro.lbl.gov/}}. 
These values of $l$ are calculated at certain energies over a range from 10 to 30000 eV. For each compound, the value of $l$
can be determined over this range, taking the influence of all the possible absorption edges of the compound into account. 
Subsequently, Equation~\ref{eq:beer_lambert} can be used to calculate the transmission $T$.
We assume that $x$ mimics an optically thin layer of dust to resemble the conditions in the ISM. 
Because $\alpha$ is independent of the depth the light travels into the material, we only need to make sure that we choose an optically
thin value of $x$. Around 1839 eV, which is the position of the Si K-edge, $l$
has a value of $~3-5 \mu$m (depending on the pre- or post edge side).
This means that if we select a thickness of 0.5 $\mu$m (a value far below the penetration depth), 
the sample becomes optically thin and in that way we can mimic the conditions in the diffuse ISM.
We now transform our absorption in arbitrary units from the laboratory data of section
3 to transmission in arbitrary units.  
This laboratory transmission spectrum can be fitted to the transmission spectrum obtained from tabulated data. 
In this way, we can determine $\alpha$ as a function of energy (or wavelength) 
in detail around the edge, since $\alpha=\frac{-\ln T}{x}$.

In order to eventually determine the absolute extinction cross-section of the silicate we need to calculate its refractive index. 
The complex refractive index $m$ is given by:
\begin{equation}
 m=n+ik,
\end{equation}
where $n$ and $k$ are the real and imaginary part of $m$ (also referred to as optical constants). 

The imaginary part of the refractive index depends on the attenuation coefficient 
$\alpha$: 
\begin{equation}
\alpha=\frac{4\pi k}{\lambda}. 
\end{equation}

Since we already obtained $\alpha$ from the laboratory data in combination with the tabulated values, the imaginary part of the refractive index 
can be derived:  
\begin{equation}
 k =  \frac{\alpha \lambda}{4 \pi}. 
\end{equation} 

\begin{figure*}
 \begin{center}
 \includegraphics[scale=0.6]{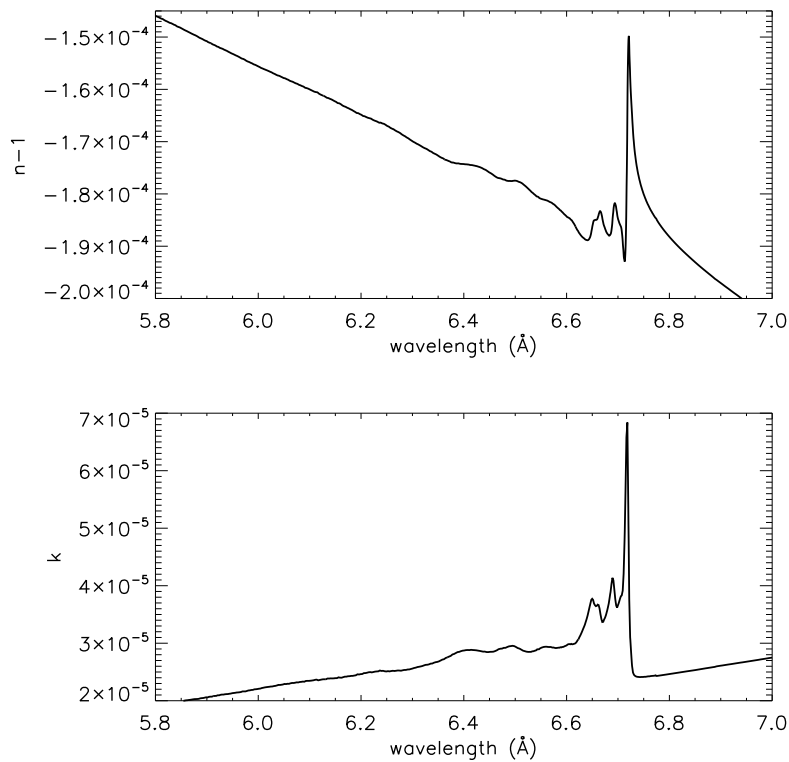}
\vspace{10pt}
 \caption{\small{The refractive index of sample 1 (olivine). 
 The upper panel shows the real of the refractive index ($n$) and the lower panel shows the imaginary part of the refractive index
 indicated by $n$ and $k$ respectively.}}
  \label{fig:real_and_imaginary}
  \end{center}
\end{figure*}

The real and imaginary part of the refractive index are not independent. 
They are related to each other by the Kramers-Kronig relations \citep{Bohren10}, in particular by:
\begin{equation}
 n(\omega)=1+\frac{2}{\pi}P\int_0^{\infty}\frac{\omega^{\prime} k(\omega^{\prime})}{\omega^{\prime2}-\omega^2}\, \mathrm{d}\omega^{\prime},\,
\end{equation}
where $\omega$ is the frequency at which the real refractive index is evaluated and $P$ indicates that the Cauchy principle value is 
to be taken. 
The real part of the refractive index can be calculated using a numerical solution of the Kramers Kronig transforms. In this paper, we use a 
numerical method using the fast Fourier transform routines (FFT) as described in \citet{Bruzzoni2002}. 
An example of the real and the imaginary part of the refractive index of sample 1 (olivine) is shown in Figure~\ref{fig:real_and_imaginary}.  

% Figure can be found in: ~/idl/idl_prog/plot_paper1/n_and_k_plot.pro

\subsection{Mie scattering calculations}

When the optical constants $n$ and $k$ are calculated, we proceed with deriving the extinction cross-section for comparison
with observational data. 
We use Mie theory to calculate the extinction efficiency ($Q_{\mathrm{ext}}(\lambda,a,\theta)$), calculated at each wavelength
($\lambda$) and particle size ($a$). 
The application of the Mie theory makes it possible to consider the contribution of both scattering and absorption to the 
cross-section. We assume a smooth dust distribution along the line of sight. 

We use the grain-size distribution of \citet{MRN77} (MRN) with a grain size interval of ($a_-,a_+$) is ($0.005 , 0.25 \mu \mathrm{m}$). 
The MRN size distribution depends on the physical and chemical state of the dust grains. 
It is described by the following equation (in the case of silicate particles): 
\begin{equation}
 n(a)da=A_{\mathrm{i}}n_{\mathrm{H}}a^{-3.5}da
\end{equation}

In this equation, $a$ is the particle size. $A$ is the normalization constant, which depends on the type of dust. 
In the case of silicate (relevant in this paper): $A_{\mathrm{sil}}=7.8\times10^{-26} \,\cent^{2.5}\,(\mathrm{H-atom})^{-1})$ (\citet{Draine84}). 
$n(a)$ is the number of grains and $n_{\mathrm{H}}$ is the number density
of H nuclei (in both atoms and molecules).  

One of the main advantages of using a MRN distribution is the simplicity of the model, which prevents the introduction of many free parameters. 
We can now calculate the extinction cross-section $C_{\mathrm{ext}}$ by applying the Mie theory (Mie 1908). 
We use the MIEV0 code \citep{Wiscombe80}, which needs $m$ and $X$, where $X=\frac{2\pi a}{\lambda}$,
as input and returns the extinction extinction cross-section $C_{\mathrm{ext}}$. 

To obtain the total scattering cross-section per wavelength unit ($\sigma_{\mathrm{ext}}(\lambda)$), 
we need to integrate over the particle size distribution. 
\begin{equation}
 \sigma_{\mathrm{ext}}(\lambda)=\int_{a_{-}}^{a_{+}} {C_{\mathrm{ext}}(a,\lambda)}n(a) da
\end{equation}

\begin{figure}
 \begin{center}
 \includegraphics[scale=0.5]{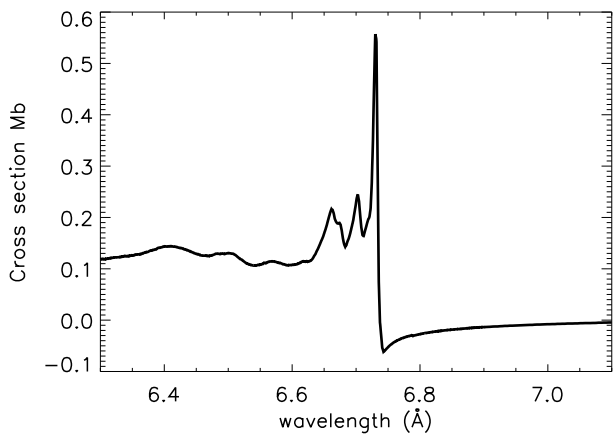}
\vspace{10pt}
 \caption{\small{Model of the Si K-edge of olivine (sample 1) as implemented in SPEX, without the continuum of the extinction profile. 
 The cross-section in Mb is given per Si-atom.}}
  \label{fig:olivine_res}
  \end{center}
\end{figure}

To make the extinction cross-sections of the compounds compatible with the Verner cross-sections used by SPEX,
we need to subtract the underlying 
continuum of the Henke tables \citep{Henke1993}. 
We remove the slope of the pre and post-edge by subtracting a smooth continuum in an energy range of 1.7-2.4 keV, which does not
contain the edge. The result for sample 1 (olivine) is shown in Figure~\ref{fig:olivine_res}.
The continua have been calculated in the same way as the edges, 
but in this case the atomic edge jump was taken out of the cross-section.
We then calculate the extinction cross-section in the same way as is described above, to obtain the continuum without the edge.
To remove the continuum of the extinction profile, we subtract the continua without edges.
This subtraction, which was done over the full energy range, 
puts cross-section of the pre-edge at zero and therefore the scattering feature before the edge obtains a negative value.
We then implement the models in SPEX. During the fitting, the continuum is naturally given by the X-ray continuum emission of the source. 

A side note has to be made that SPEX fits the edges to the Verner tables instead of Henke. 
This means that there is a small discrepancy between the two parametrizations of approximately 2-5 percent,
around the silicon edge (J. Wilms, private communications). 
This is well within the limit of precision 
that we can achieve from observations. 

\section{GX~5-1}

As a first test of these new samples, we apply the new models to the source GX~5-1, which serves as a background
source to observe the intervening gas and dust along the line of sight. 
GX~5-1 is a bright low-mass X-ray binary at $(l,b)=(5.077,-1.019)$. \citet{Christian97} estimated distance 
to GX~5-1 to be 9 kpc. However, this distance may be regarded as an upper limit and the actual distance could be up to $30\%$ (or 2.7 kpc) less,
because the luminosity derived from the disk model used in their analysis exceeds the Eddington limit for accretion of either
hydrogen or helium.
GX~5-1 has been observed multiple times, for instance by \textit{Einstein} \citep{Christian97,Einstein79}, as well as ROSAT using the Position
Sensitive Proportional Counter \citep{Predehl95}, 
and there are several observations by \textit{Chandra}. 
In this work we use observations of the high-resolution spectrum of GX~5-1 collected by
the HETG instrument on board \textit{Chandra}. 
The Galactic coordinates combined with the distance indicate that GX5-1 is located near the Galactic center
(assuming a distance towards the Galactic center of 8.5 kpc). 
Due to the uncertainty in the distance of GX~5-1, the source may either be in front, behind of embedded 
in the Galactic center region \citep{Smith06}.
The proximity to this region enables us to probe the dust and gas along one of the densest sight lines of the Galaxy.
From CO emission observations towards the line of sight of GX~5-1 (observed by \citet{Dame01}), 
\citet{Smith06} concluded that there are three 
dense regions along the line of sight, namely the 3 kpc spiral arm, the Giant Molecular Ring and the Galactic center. 
The three regions are at distances of 5.1, 4.7, and 8.5 kpc respectively, \citep{Smith06}.  
The Giant Molecular Ring is assumed to be a region with a high molecular cloud density, which can be found approximately half  
way between the Sun and the Galactic center.
All these regions may provide a contribution to the observed dust along the line of sight.
The observation in the analysis of this paper therefore consists of a mixture of dust and gas in these regions.

\section{Data analysis of GX~5-1}

The spectrum of GX~5-1 has been measured by the HETG instrument of the Chandra space telescope. This instrument contains two gratings:
HEG and MEG with a resolution of $0.012\,\AA$ and $0.023\,\AA$ (full width at half maximum), respectively \citep{HETG2005}. 
The energy resolution of our models is therefore well within the resolution of the Chandra Space Telescope HETG detector.
There are multiple observations of GX~5-1 available in the Chandra Transmission Gratings Catalog and Archive
\footnote{\tiny{\burl{http://tgcat.mit.edu/}}}, 
but only the observations in TE mode could be used to observe the Si K-edge (OBSID 716). This edge is not readily visible in the CC-mode, 
where it is filled up by the bright scattering halo radiation of the source.
The edge has a slight smear as well as different optical depth. 
This effect is particularly evident in the CC mode, where the two arms of the grating 
are now compressed into one dimension, together with the scattering halo image
(N. Schulz, private comm.).\footnote{\tiny{\burl{http://cxc.harvard.edu/cal/Acis/Cal_prods/ccmode/ccmode_final_doc03.pdf}}}

GX~5-1 is the second brightest persistent X-ray source after the Crab Nebula \citep{Smith06}   . 
Due to the brightness of the source (with a flux of $F_{0.5-2 \mathrm{keV}}=4.7\pm0.8\times10^{-10}\,\mathrm{erg}\,\mathrm{cm}^{-2}\,\mathrm{s}^{-1}$ and
$F_{2-10 \mathrm{keV}}=2.1\pm0.6\times10^{-8}\mathrm{erg}\, \mathrm{cm}^{-2}\,\mathrm{s}^{-1}$~\ref{table:model_columnsdens}), 
the Chandra observation suffers from pile-up. This 
effect is dramatic in the MEG data, but also has an effect on the HEG grating. 
We, therefore, cannot use the MEG grating and need to ignore the HEG data below $4.0\,\AA$.
However, in this observation, GX~5-1 was also observed using both a short exposure of 0.2385 ks and 
a long exposure of 8.9123 ks. This is not a standard mode, but 
was especially constructed to evaluate the pile-up in the long exposure. The spectrum with an exposure time of 0.2385 ks 
does not suffer from pile-up due to the short exposure time per frame in the TE mode and 
can therefore be used to determine both the continuum and the hydrogen column density of the source. 

Figure~\ref{fig:continuum} shows the broad band spectrum of GX~5-1. The spectrum shows the presence of a strong Si K-edge
at $6.7\AA$  superimposed on a strongly rising continuum towards shorter wavelength.
The Si K-edge shows a strong transition with clear absorption structure at shorter wavelength, which is the tell-tale
signature of solid-state absorption and an anomalous dispersion peak at the long wavelength, which reveals the 
presence of large grains~\citep{Hulst1958}.

\subsection{Continuum and neutral absorption}

Before we can calculate the mixture of dust that best fits the data of GX~5-1, we need to determine the column density of hydrogen ($N_{\mathrm{H}}$)
towards GX~5-1 and the underlying continuum of the source. 
Earlier research of \citet{Predehl95} using data from the ROSAT satellite shows that the value of the column density
ranges between $~2.78 \,\mathrm{and}\, 3.48 \times10^{22}\,\cent^{-2}$ depending on the continuum model. 
The disadvantage of the ROSAT data is that it only covers the lower-energy side (with 2 keV as the highest energy available \citep{Truemper82})
of the spectrum and therefore prevents the fitting of the hard X-ray energy side. This made it hard to predict which model would fit the spectrum
the best and in that way influenced the value of $N_{\mathrm{H}}$. 
Other more recent measurements of the value of the column density were taken by \citet{Ueda05}; $2.8\times10^{22}\,\cent^{-2}$,
based on fits on the same Chandra HETG data used in this analysis and 
\citet{Asai00}; $3.07\pm0.04 \times10^{22}\,\cent^{-2} $ (using ASCA archival data). 

\begin{figure}
 \begin{center}
 \includegraphics[scale=0.5]{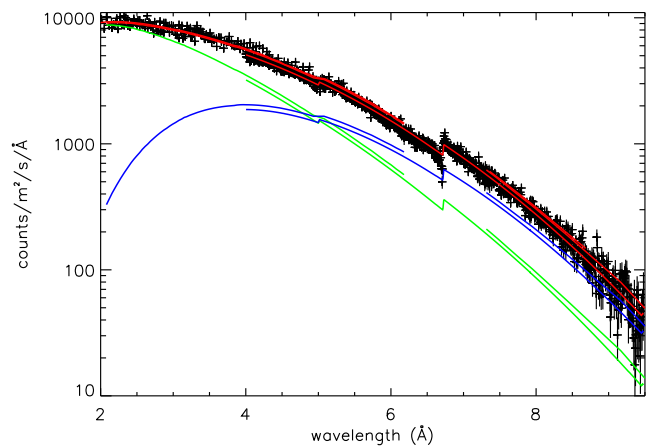}
 \vspace{10pt}
 \caption{\small{The continuum of GX~5-1. 
 Data from the HEG short exposure, MEG short exposure and HEG long exposure (OBSID 716) were used to fit the continuum.
 The resulting fit to each of the three data sets is shown by a red line. The model consists of two absorbed 
 black bodies. The two black bodies in the model are shown in figure by the green
 and blue lines. For clarity, only the black body models of the HEG grating (both long and short exposure) are shown and
 the MEG data has been omitted in the figure, 
 therefore each black body model shows two curves.}}
 \label{fig:continuum}
  \end{center}
\end{figure}

The short Chandra HETG exposure does not suffer from pile-up and contains both the soft and the hard X-ray tail of the spectrum,
therefore it can be used to derive the continuum and the $N_{\mathrm{H}}$ for this analysis. 
The spectrum is best modelled by two black body curves using the bb model in SPEX.
The spectrum is absorbed by a cold absorbing neutral gas model, simulated
by the HOT model in SPEX. The temperature of this gas is frozen to a value of $kT=5\times10^{-4}\,\mathrm{keV}$, in
order to mimic a neutral cold gas. HEG and MEG spectra of the short exposure are fitted simultaneously with this model. 
The best fit for the continuum and the $N_{\mathrm{H}}$ is shown in Figure~\ref{fig:continuum} and
Table~\ref{table:model_columnsdens}. 
It is necessary to constrain the $N_{\mathrm{H}}$ well on the small wavelength ranges that are used in the following part of the analysis. In
this way, the continuum is frozen and cannot affect the results of the dust measurements (see section 6.2). 
Therefore, we use the HEG data of the long exposure to further constrain the column density. 
In the short exposure, we ignore the data close to the edge (6.2 - 7.2 $\AA$), because the long exposure provides a much
more accurate measurement of this part of the spectrum. In this way, the fit will not be biased by the lower signal-to-noise 
in the short exposure in this region.  
When we fit the model to the data, we obtain a good fit to the data with $C^2/\nu=1.16$. For now, we only take cold gas into account
to fit the continuum spectrum. 
This resulted in a column density of $3.40\pm0.1\times10^{22}\,\mathrm{cm}^{-2}$ (see Figure~\ref{fig:continuum} and 
Table~\ref{table:model_columnsdens}).
All the fits in this paper generated by SPEX are using C-statistics \citep{Cash79} as an alternative to $\chi^2$-statistics. 
C-statistics may be used regardless of the number of counts per bin, so in this way we can use bins with a low count rate in the spectral fitting. 
\footnote{\tiny{See for an overview about C-statistics the SPEX manual: \burl{https://www.sron.nl/files/HEA/SPEX/manuals/manual.pdf}
and \burl{http://heasarc.gsfc.nasa.gov/lheasoft/xanadu/xspec/manual/XSappendixStatistics.html}}}
Errors given on parameters are $1\,\sigma$ errors. 

% Figure can be found in: ~/idl/idl_prog/spex_prog/all_over_again/all_with_newmodels/make_cont_plot.pro

\begin{table}
\caption{Broad band modelling of the source using HEG and MEG data from Chandra HETG} % title of Table
\label{table:model_columnsdens} % is used to refer this table in the text
\centering % used for centering table
\begin{tabular}{l l l } % centered columns (4 columns)
\hline\hline\noalign{\smallskip} % inserts double horizontal lines
% \rule{0pt}{2.3ex}
     $N_\mathrm{H}$&$3.4\pm0.1\times10^{22}$ & $\cent^{-2}$\\ 
     $T_{bb1}$& $0.59\pm0.02$ & keV \\
     $T_{bb2}$& $1.44\pm0.05 $ & keV \\
     $F_{0.5-2 \mathrm{keV}}$& $4.7\pm0.8\times10^{-10}$ & $ \,\mathrm{erg}\, \mathrm{cm}^{-2}\,\mathrm{s}^{-1}$  \\
     $F_{2-10 \mathrm{keV}}$& $2.1\pm0.6\times10^{-8}$& $\mathrm{erg}\, \mathrm{cm}^{-2}\,\mathrm{s}^{-1}$  \\
     $C^2/\nu$& 1169/1005 & \\ 
\hline %inserts single line
\end{tabular}
\end{table}

\subsection{Fit to Chandra ACIS HETG data of the silicon edge}

After obtaining the $N_\mathrm{H}$ value, we fit the dust models of the six dust samples to the Chandra HETG data.  
The shape of the continuum is fixed for now, to avoid any dependence of the fit on the continuum. 
The column density used for this fit is the $N_\mathrm{H}$ derived in section 6.1 and can vary in a range of $1\sigma$ from this 
value. To rule out any other dependence, we only use the data 
in a range around the edge: $6-9\,\AA$. In this way we include the more extended XAFS features as well as part of the continuum
in order to fit the pre and post edge to the data.
The depletion values and ranges that were assumed for the cold gas component are listed in Table~\ref{table:depletions}. 
Furthermore, SPEX uses protosolar abundances for the gas phase that are taken from \citet{Lodders09}. 

\begin{table}
\caption{Depletion ranges used in the spectral fitting} % title of Table
\label{table:depletions} % is used to refer this table in the text
\centering % used for centering table
\begin{tabular}{l c} % centered columns (4 columns)
\hline\hline % inserts double horizontal lines
Element & Depletion range  \\ % table heading
\hline\noalign{\smallskip} % inserts single horizontal line
% \rule{0pt}{2.3ex}
Silicon & $0.8-0.97$  \\ % inserting body of the table
Iron & $0.7-0.97$  \\
Magnesium & $0.9-0.97$  \\
Oxygen & $0.2-0.4$  \\
\hline %inserts single line
\end{tabular}
\tablefoot{Depletion ranges in this table are based on depletion values from \citet{Wilms00}, 
\citet{Costantini2012} and \citet{Jenkins09}.}
\end{table}

The SPEX routine AMOL can fit a dust mixture consisting of four different types of dust at the same time. Therefore,
we test all possible configurations of the dust species and compare all the outcomes. We follow the same method as 
described in \citet{Costantini2012}, where the total number of fits $n$ is given by $n={n_\mathrm{edge}}!(4!({n_\mathrm{edge}}-4)!)$
and ${n_\mathrm{edge}}$ is the number of available edge profiles.

\begin{figure*}
 \begin{center}
 \includegraphics[scale=0.7]{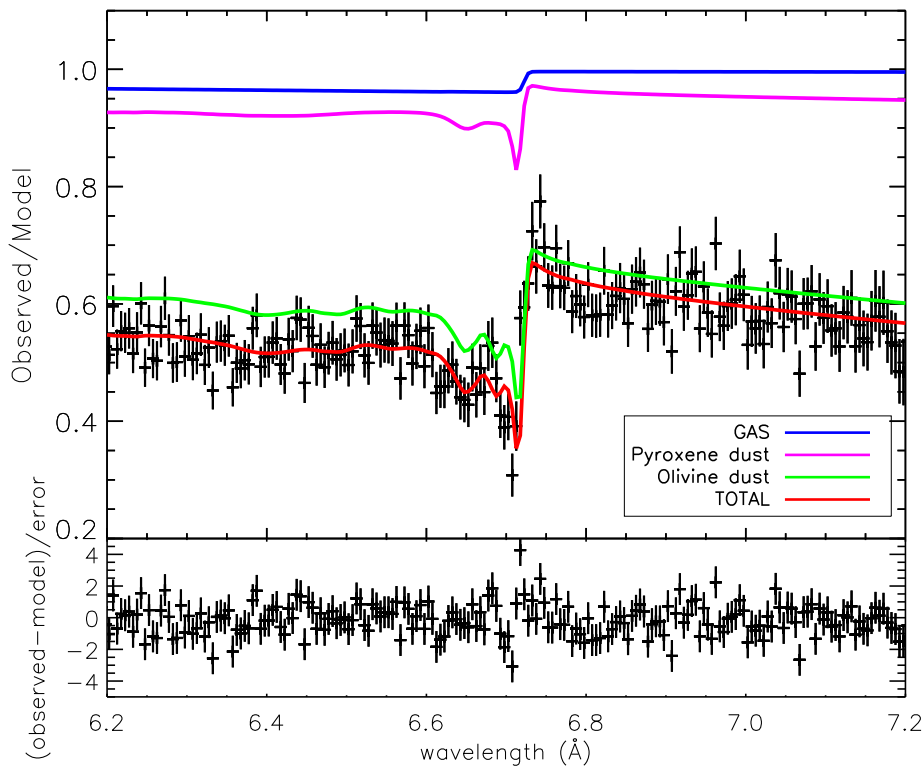}
 \vspace{2pt}
 \caption{\small{The fit to the spectrum of GX~5-1 is shown in red. The contribution of the continuum is divided out. 
 The other lines shows the contribution of the absorbing components to the transmission. 
 The purple line shows the contribution of pyroxene (sample 5), the green line the contribution of olivine (sample 1)
 and the blue line the contribution of gas. 
 The lower pannel shows the model residuals of the fit in terms of the standard deviation, $\sigma$.}}
 \label{fig:fit}
  \end{center}
\end{figure*}

In Figure~\ref{fig:fit} we show the results of a fit of the SPEX model to the observed spectrum of GX~5-1. 
The best fit is shown by the red line. The mixture that fits the data best consists mainly of crystalline olivine (sample 1, green line) 
and a smaller contribution of amorphous pyroxene (sample 5, purple line) and neutral gas (blue line).

% Figure can be found in: ~/idl/idl_prog/spex_prog/all_over_again/all_with_newmodels/second_ref/new_bestfitplot_res/plot_best_fit.pro

The reduced $C^2$ value of the best fit is 1.05. We find that most dust mixtures that contain olivine fit the edge well 
and when pyroxene, with a high concentration of iron, is added to the fit, we obtain even better fits.  
When olivine is left out of the fit, the reduced $C^2$ values increase from
1.05 to values of approximately 2. 
From our fitting procedure, we conclude that the dust mixture consists of $86\pm7\%$ olivine dust and $14\pm2\%$ pyroxene dust.

We calculated the dust abundances of silicon, oxygen, magnesium and iron to be respectively
$4.0\pm0.3\times10^{-5}\mathrm{per}\,\mathrm{H}\,\mathrm{atom}$,
$17\pm1\times10^{-5} \,\mathrm{per}\,\mathrm{H}\,\mathrm{atom}$, $6.1\pm0.3\times10^{-5} \,\mathrm{per}\,\mathrm{H}\,\mathrm{atom}$,
and $1.7\pm0.1\times10^{-5} \,\mathrm{per}\,\mathrm{H}\,\mathrm{atom}$. 
Unfortunately, the depletion hits the higher limit of the ranges set in Table \ref{table:depletions}. 
These ranges were set in order to keep the fit within reasonable depletion values.
For this reason, we can only give upper limits to the depletion values we found, which are for O: $<0.23$, 
for Mg: $<0.97$, for Si: $<0.87$ and for Fe: $<0.76$. 
Table~\ref{table:abundances_depletions} lists the total column density, the depletion values, dust abundances, total abundances (including both 
gas and dust), and solar abundances for all the elements mentioned above.
The total abundances can only be given as lower limits, since we can only put a 
lower limit on the gas abundance due to the upper limits on the depletion values.  
From our best fitting dust mixture, the total abundance of silicon along the line of sight towards GX~5-1 can be calculated 
using the column density of the best fit and the total amount of silicon atoms in both gas and solid phase. 
The resulting abundance is: $>4.4\times10^{-5} \,\mathrm{per}\,\mathrm{H}\,\mathrm{atom}$. 
The gas to dust ratio of silicon is 0.22.

\begin{table*}
\caption{Abundances and depletions} % title of Table
\label{table:abundances_depletions} % is used to refer this table in the text
\centering % used for centering table
\begin{tabular}{l c c c c c} % centered columns (4 columns)
\hline\hline % inserts double horizontal lines
Element& $N^{\mathrm{tot}}$ & depletion & $A_{Z}$ & $A_{Z}^{\mathrm{dust}}$ & $A_Z/A_{\sun}$\\ % table heading
& & & $(10^{-5}\,\mathrm{per}\,\mathrm{H}\,\mathrm{atom})$& $(10^{-5}\,\mathrm{per}\,\mathrm{H}\,\mathrm{atom})$& \\
\hline\noalign{\smallskip} % inserts single horizontal line
Silicon &$>1.4$& $<0.87$ &$>4.4$ & $4.0\pm0.3$ & $>1.14$\\ % inserting body of the table
Iron &$>0.7$& $<0.76$ & $>2.5$ & $1.7\pm0.1$ & $>0.79$\\
Magnesium &$>2.1$& $<0.97$ & $>6.2$ & $6.1\pm0.3$ & $>1.6$\\
Oxygen &$>23$& $<0.23$ & $>64$& $17\pm1$ & $>1.06$ \\
\hline %inserts single line
\end{tabular}
\tablefoot{Abundances are indicated by $A_{Z}$. Solar abundances are taken from \citet{Lodders09}. Total column densities (gas and dust) 
$N^{\mathrm{tot}}$ are in units of $10^{18}\mathrm{cm}^{-2}$.}
\end{table*}

\subsection{Hot ionized gas on the line of sight in the Si K-edge region?}
Hot ionized gas along the line of sight might influence the silicon edge. In the energy range of the edge,
we may observe absorption lines of this hot gas that may influence the shape of the silicon absorption edge. 
It may also be possible that this hot gas is intrinsic to the source.
To be certain that this is not 
the case, we fitted the edge again as described above, but this time we add a slab of hot gas along the line of sight to our model.
The lower and upper limits of the gas temperature are based upon the ionization fractions of neon by \citet{Yao05} and \citet{Yao06}. 
They observed hot gas in the ISM and
compare the observed lines to ionization fractions for O, Ne, and Fe to determine the temperature of the gas. 
In this paper, we use the NeIX line to set the lower and upper limit of the hot gas (0.08 - 2.7 keV). 
At these temperatures of the hot gas, helium-like transitions of Si can occur at 6.69 and 6.65 $\AA$.
These lines can create additional
features in the Si K-edge, so we fit the edge again with SPEX including an extra HOT model to model the hot gas.  
We do not detect any ionized gas along the line of sight towards GX~5-1. The gas temperature of the hot model hits the set
limit of 0.08 keV, which is too cold to form any absorption lines near the silicon K-edge.  
Any ionized gas is therefore not likely to contaminate the Si K-edge in this data set. 

\section{Discussion}

\subsection{Abundances towards GX~5-1}

\subsubsection{Si abundance}

In section 6.2 we find that the abundance of silicon in dust is  $4.0\pm0.3\times10^{-5}\,\mathrm{per}\,\mathrm{H}\,\mathrm{atom}$. 
We can compare this result to observations at infrared wavelengths in the solar neighborhood and 
towards the Galactic center. The 10 and 20$\,\mu\mathrm{m}$ lines of the Si-O bending and stretching modes were observed
in order to measure the silicon abundances. 
We derived these abundances using the results from \citet{Aitken84}, \citet{Roche85}, and \citet{Tielens96}.
The silicon abundance was derived in two different regions of the Galaxy, namely the local solar neighborhood and 
a region close to the Galactic center. In the local
solar neighborhood towards sight lines of bright nearby Wolf-Rayet stars the abundance of silicon in dust can be derived, resulting
in a value of
$5.2\pm1.8\times10^{-5}$ per H atom \citep{Roche84,Tielens96}.
Measuring the Si abundance towards the Galactic
center from the $10\,\mu\mathrm{m}$ absorption may be challenging.
The abundances depend on an estimate of the visual extinction ($A_V$)
derived from the $N_H/A_V$ ratio for the local Solar neighborhood from UV studies of the
atomic and molecular hydrogen column densities \citep{Bohlin78}, and this procedure 
may be more uncertain towards the Galactic center \citep{Tielens96}.     
The silicon abundance may therefore suffer from additional uncertainty.
The same analysis has also been carried out for sight lines towards the Galactic center \citep{Tielens96, Aitken84}. 
Towards the sight line of a cluster of compact infrared sources 2 pc away from the Galactic center \citep{Roche85},
the silicon abundance in dust was derived, resulting in a value of 
$3.0\pm1.8\times10^{-5}\,\mathrm{per}\,\mathrm{H}\,\mathrm{atom}$. 
This is almost half of the value of the dust abundance measured in the local solar neighborhood. 
This discrepancy is not well understood. An explanation could be that the difference in 
abundance is caused by presence of large particles (grain sizes $>3\,\mu\mathrm{m}$) near the Galactic center,
which, however, is difficult to observe in the infrared \citep{Tielens96}. 

The abundance of silicon in dust found in section 6.2, 
falls in between the two results for the local ISM and the Galactic center
obtained in the infrared. 
When we consider the total abundance (including the contribution from gas), this increases the 
total abundance to $>4.4\times10^{-5}$ per H atom,
which would correspond more with values found in the local solar neighborhood than those of the Galactic center. 
When we compare the total abundance of silicon to the protosolar
abundance from \citet{Lodders09}, we find only a small deviation from the solar abundance, namely: $A_Z/A_{\sun}>1.14$. 
Furthermore, since we probe the inner regions of the Galaxy, it is not unrealistic to encounter a total abundance of silicon
larger than solar along the line of sight.  

Another comparison can be made using the observations towards the low-mass X-ray binaries 4U 1820-30 \citep{Costantini2012}
and X Per \citep{Valencic13}. In contrast to GX~5-1, these lines of sight probe the diffuse ISM. 
Towards 4U 1820-30 the abundance is $4.8^{0.8}_{-0.5}\times10^{-5}\,\mathrm{per}\,\mathrm{H}\,\mathrm{atom}$ 
and towards X Per is $3.6\pm0.5\times10^{-5}\,\mathrm{per}\,\mathrm{H}\,\mathrm{atom}$. The lower limit 
of the Si abundance found in our analysis does, therefore, agree with the results of the diffuse ISM. 
The actual value of the Si abundance might indeed be higher than the average Si abundance in the diffuse ISM, 
but we need upper limits to the abundances to confirm this by obtaining better estimates on the depletion values. This 
may be obtained by releasing the range on the silicon depletion, which is discussed in section 7.1.2.

\subsubsection{Depletion of silicon}

\begin{figure}
 \begin{center}
 \includegraphics[scale=0.5]{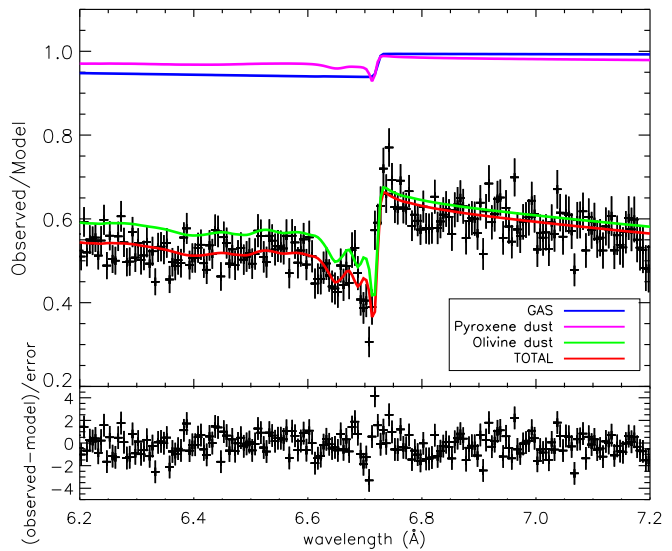}
 \vspace{10pt}
 \caption{\small{The upper panel shows a fit of the Si K-edge, where the depletion of silicon is left as a free parameter without boundary values. 
 The best fitting mixture
 consist out of the same compounds as the fit in section 6.2: sample 1 (crystalline olivine) and sample 5 (amorphous pyroxene). 
 The lower panel shows the model residuals of the fit in terms of the standard deviation, $\sigma$.}}
 \label{fig:Unconstrained_si}
  \end{center}
\end{figure}

Since the depletion value of silicon reached the lower limit, we also show the results of a fit without restrictions on the
silicon depletion values. These results are shown in Table~\ref{table:abundances_depletions_free} and Figure~\ref{fig:Unconstrained_si}.
By releasing the range on the silicon depletion, we get a better impression of the total abundance of silicon. 

\begin{table*}
\caption{Unconstrained silicon: abundances and depletions} % title of Table
\label{table:abundances_depletions_free} % is used to refer this table in the text
\centering % used for centering table
\begin{tabular}{l c c c c c} % centered columns (4 columns)
\hline\hline % inserts double horizontal lines
Element& $N^{\mathrm{tot}}$ & Depletion & $A_{Z}$ & $A_{Z}^{\mathrm{dust}}$ & $A_Z/A_{\sun}$\\ % table heading
& & & $(10^{-5}\,\mathrm{per}\,\mathrm{H}\,\mathrm{atom})$& $(10^{-5}\,\mathrm{per}\,\mathrm{H}\,\mathrm{atom})$& \\
\hline\noalign{\smallskip} % inserts single horizontal line
Silicon &$1.8\pm0.3$& $0.68\pm0.12$ &$5.0\pm0.5 $ & $3.8\pm0.5$ & $1.30\pm0.12$\\ % inserting body of the table
Iron &$>0.7$& $<0.75$ & $>2.5$ & $1.7\pm0.2$ & $>0.76$\\
Magnesium &$>2.1$& $<0.97$ & $>6.1$ & $6.3\pm0.5$ & $>1.6$\\
Oxygen &$>23$& $<0.22$ & $>0.63$& $16\pm2$ & $>1.03$ \\
\hline %inserts single line
\end{tabular}
\tablefoot{Same as Table~\ref{table:abundances_depletions}, 
but depletion of silicon is left as a free parameter without boundary values. 
Abundances are indicated by $A_{Z}$. Solar abundances are taken from \citet{Lodders09}. Total column densities (gas and dust) 
$N^{\mathrm{tot}}$ are in units of $10^{18}\mathrm{cm}^{-2}$.}
\end{table*}

Since we are currently analyzing the K-edge of silicon, 
we are able to directly measure the abundance of depletion in both gas and dust of this element only. 
The abundances of other elements are derived indirectly from the model and their depletion values should thus be kept in a limited range. 
We performed the fit again, leaving the depletion of Si free to vary without predefined boundaries.
In this case, we obtain a depletion value of $0.68\pm0.12$. 
This value is within the 1 sigma error consistent with the value in
Table~\ref{table:abundances_depletions}. 
The best fit in Figure~\ref{fig:Unconstrained_si} shows the same samples as the best fit in section 6.2 (namely sample 1 olivine
and sample 5 pyroxene), but there is a relatively larger 
contribution of gas, at the expense of the amorphous pyroxene contribution.
As a consequence, the dust abundance value is reduced. 
Both $A_Z$ and $A_Z/A_{\sun}$ are now constrained. $A_Z/A_{\sun}$ seems to show a clearer overabundance of Si. 
This is expected in environments close to the Galactic center.
The value of the total silicon abundance, $A_{Z}=5.0\pm0.5\times10^{-5}\,\mathrm{per}\,\mathrm{H}\,\mathrm{atom}$,
is comparable with values found in the local solar neighborhood.
The differences between $N^{\mathrm{tot}}$, the depletion, $A_{Z}$, $A_{Z}^{\mathrm{dust}}$, and $A_Z/A_{\sun}$
in Table~\ref{table:abundances_depletions} and Table~\ref{table:abundances_depletions_free} are small
in the case of the other elements: Fe, Mg, and O.
This is expected, because the range on the depletion ranges of these elements were kept the same in both fits. 
The number of fit parameters did not change, nor did the quality of the fit. Therefore, the reduced $C^2$ value of the fit remains at 1.05.
However, since the depletion of silicon in dense environments is expected to be higher than $0.68\pm0.12$, as is shown in previous studies 
\citep{Wilms00,Jenkins09,Costantini2012}, we conservatively keep the silicon 
depletion constrained by the limits in Table~\ref{table:abundances_depletions} in further analysis of the edge. 
% In future studies, the depletion of silicon may 
% be better constrained by simultaneously fitting the magnesium K-edge, which imposes additional constraints on the dust abundance.} 

% Figure can be found in: ~/idl/idl_prog/spex_prog/all_over_again/all_with_newmodels/second_ref/unconstrained_mod/best_fit_plot/plot_this_stuf.pro

\subsubsection{Other abundances: oxygen, magnesium and iron}

The abundances of the other elements are listed in Table~\ref{table:abundances_depletions}. 
Silicon and oxygen show abundances comparable to the solar values of~\citet{Lodders09}. Magnesium and iron 
show a deviation from protosolar abundances.
The abundances in the Galaxy follow a gradient of increasing 
abundance towards the plane of the Galaxy. In the inner regions of the Galaxy, the abundances of the elements are expected 
to be supersolar (see section 1).
Most of the lower limits on the abundances correspond to values found in studies of the diffuse ISM
\citep{Pinto10,Costantini2012,Valencic13}. Since we can only show lower limits, we expect the true abundance to be higher, 
which would correspond with the expected increase of abundances toward the region around the Galactic center.
The abundance of magnesium is supersolar, whereas the iron abundance in our result is subsolar. The missing iron can be 
present in other forms of dust. A possibility could be that iron is included in metallic form in GEMS \citep{Bradley94}.

\subsection{Comparison to iron poor, amorphous, and crystalline dust}
\begin{figure}
 \begin{center}
 \includegraphics[scale=0.5, right]{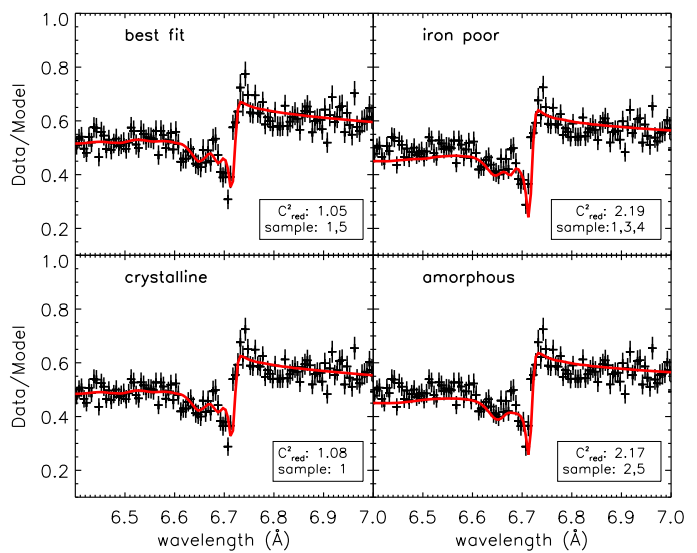}
 \caption{\small{Comparison of the best fit (upper left window) with iron poor dust mixtures, strictly crystalline mixtures and amorphous mixtures.}}
 \label{fig:comparisonfit}
  \end{center}
\end{figure}

An insightful way to compare the best fitting mixture to all the other possible mixtures is by comparing 
the best fitting mixtures with the two distinct variations
in the samples: iron-rich/iron-poor and crystalline/amorphous. We select three different types of mixtures that we 
compare to the best fitting mixture, namely an iron-poor mixture (since the best-fitting mixture is iron-rich), 
a mixture of amorphous compounds and a mixture of crystalline compounds. 
The results are shown in Figure~\ref{fig:comparisonfit}. 
  
% can be found in  ~/idl/idl_prog/spex_prog/all_over_again/all_with_newmodels/compare_again/make_total_plot.pro

The fit that resembles the best fit the most, is the fit containing only crystalline compounds. This fit consists only of
olivine, which is the compound that is dominating the best fit of section 6.2. The fit improves when amorphous pyroxene
is added. The different compounds show a strong variation in the 
absorption at wavelengths of less than $6.6\AA$ relative to the K-edge at $6.7\AA$
Fig~\ref{fig:all_models}. This difference causes the poor fits for amorphous or
iron samples (Fig~\ref{fig:comparisonfit}).
The resulting fits in Figure~\ref{fig:comparisonfit} strongly depend on the amount of silicon atoms available in the solid state,
which is restricted by the column density $N_{\mathrm{H}}$.
As can be observed in Fig~\ref{fig:all_models} not every edge of the samples is equally deep. 
Models that produce deep absorption features around the edge, 
demand the presence of a certain amount of silicon in dust present along the line of sight. The amount of silicon present
constrains the best fitting model in this way. The result of fitting a
sample to the edge that requires a larger amount of silicon along the line of sight can be observed in the iron poor and 
amorphous windows of Fig~\ref{fig:comparisonfit}. Here the edge of the models is too deep and the post edge (which is 
here the part below $6.6\AA$) never recovers to fit the data in this part of the plot. The amount of Si along the line of sight can 
be enhanced if the $N_{\mathrm{H}}$ increases. When we enhance $N_{\mathrm{H}}$ to values $>5\times10^{22}\,\mathrm{cm}^{-2}$ 
the iron poor and amorphous models start to fit the edge, but such a $N_{\mathrm{H}}$ is too high for GX~5-1.   
A well-determined column density of hydrogen is, for this reason, of great importance.

The best fit in section 6.2 shows that we detect $86\pm7\}$ of crystalline olivine dust and only $14\pm2\%$
of amorphous pyroxene dust. 
This result gives us a relatively high crystallinity, that is, 
$79-93\%$ of the total amount of dust has a crystalline structure. 
This somewhat surprising result may be in line with recent results from 
the Stardust mission \citep{Westphal14} and detection of dust in external galaxies
\citep{Spoon2006,Kemper2011}, which suggests that 
crystalline dust may be more abundant than expected~\citep{Kemper04}; see section 1.
Limitations on this result are addressed in section 7.4. 

\subsection{Comparison with dust compositions along other sight lines}
We can compare the best fitting dust mixture to results from studies of dust in the infrared and X-ray studies of dust
in the diffuse ISM. 
The bending and stretching modes of silicon in the infrared are often attributed to a mixture of enstatite and olivine 
\citep{Kemper04,Chiar06}. More detailed studies on the dust composition can be acquired from X-ray studies. 
Studies of the diffuse ISM show variation of the composition of the ISM along different lines of sight
\citep{Lee09,Pinto10, Costantini2012, Valencic13}. 
From studies of the Fe L and O K-edge, \citet{Pinto10} concluded that the dust composition of interstellar dust
is probably chemically inhomogeneous, since they found indications for iron rich silicates toward X-ray binary GS 1826-238. 
\citet{Lee09}, on the contrary, find that Fe oxides provide a better fit to the data of Cyg X-1. 
\citet{Costantini2012} concluded from an analysis of the oxygen K-edge and the iron L-edge
of the source 4U 1820-30, that GEMS could possibly be present along this line of sight.
Their best fitting mixture, is a mixture of enstatite and metallic iron,
which suggests that the dust in this region mostly consists of enstatite with metallic inclusions.  

Along the line of sight toward X Per a dust mixture of $\mathrm{MgSiO}_3$ and 
iron bearing silicates such as $\mathrm{Mg}_{1.6}\mathrm{Fe}_{0.4}\mathrm{SiO}_{4}$
provide a good fit to the observed spectrum \citep{Valencic13}. They conclude that although $\mathrm{MgSiO}_3$ is the dominating 
compound, the fit improves when iron-bearing silicates are included. 
The best fitting dust mixture in this study consists of the compounds that contain the largest amount of iron available in 
our set of compounds. Compared to the 
amount of magnesium present in these samples however, we can still consider this dust mixture as relatively iron poor. 
This is also reflected in the abundances of iron and magnesium of the best fit. 

\subsection{Limiting factors in the analysis of the Si K-edge}

The modeling of a photo-electric edge such as the  Si K-edge involves many components, 
which may influence the resulting shape of the edge. This requires a careful analysis of the resulting
best-fit. This paragraph provides more insight into the quality of the fits and the limitations that arise in the analysis. 

The quality of the fit depends in the first place on the resolution of the instrument.
The observations used in this analysis were observed with the HEG grating of the \textit{Chandra} X-ray telescope. 
This grating provides us with the best resolution currently available close to the Si K-edge. 
The bright source provides us with a high flux near the Si K-edge and the edge falls well within the range of the grating, 
where the effective area of the instrument is large.
The minimum S/N per bin near the Si K-edge is 20, which is sufficient for the study.

Additionally, the accuracy of the results depends on the degrees of freedom of the fit 
and, of course, on the completeness of the dust models.
SPEX allows a maximum input of four different dust components in the same fit.
If we examine the output of the fits for all the possible dust mixtures, most of the resulting fits contain two 
dust components with negligible contribution from the other two possible components. In almost all the 
fits, one of the dust components dominates. The contribution of the second component is, in most cases, no more than
$20\%$ and in two cases there is a contribution of a third component, which is never more than $1-2\%$. 
When we analyze the fits that are within 3 sigma of the best-fit values presented in section 6.2, the dominating component in
these fits is crystalline olivine. Interestingly, all the dust mixtures where olivine is not included do not fall within 3 sigma of 
our best fit and can be ruled out.

Our set of models is sufficient for a pilot study of the Si K-edge, but should be expanded for further analysis. 
The main limiting factor in this analysis is the absence of an amorphous olivine model as a counterpart to the crystalline olivine model
(See section 3.1). 
We note that the crystalline counterpart cannot be taken as a proxy for the amorphous one.
As an example, see the profiles of pyroxenes in amorphous and crystalline form in Figure~\ref{fig:crystal_and_amorphous}. 
The sharp features of the XAFS in the observation of GX~5-1 are indicative of a crystalline structure 
(Figure~\ref{fig:crystal_and_amorphous}). However,
if the shape of the amorphous olivine would follow that of a crystalline counterpart albeit less sharp, the fit may 
allow the presence of more amorphous olivine. At the moment we are unable to test this (see the discussion in section 3.1). 
Therefore, we conclude that the interesting possibility of a high fraction of crystalline dust along this line of sight 
must be further verified.

Another limitation in our analysis could be the Mg:Fe ratio in our samples. 
The resulting abundances and depletion values in our analysis depend strongly on the best-fitting dust mixture.
The silicates in our sample are mainly olivines and pyroxenes with different Mg:Fe 
ratios. All the samples are relatively iron poor with a maximum Fe contribution of Mg/(Mg+Fe) of 0.6 and although
there are indications that interstellar and circumstellar may indeed be iron poor
\citep{Molster202,Molster102,Molster302,Chiar06,Blommaert14}, 
it would be beneficial to include silicates with a higher iron content. 

In order to better constrain the depletion values, it is necessary to expand the fit to a broader wavelength range
and incorporate multiple edges in the analysis depending on the observed $N_\mathrm{H}$.  
In the case of GX~5-1 iron cannot be directly observed. The $N_\mathrm{H}$ is not high enough to imprint a significant Fe K-edge.
On the lower energy side, interstellar absorption suppresses the Fe L-edges at $\sim\,0.7$ keV and the O K-edge at 0.543 keV.
The iron abundance in silicates can therefore only be inferred from
the best fitting dust mixture. 
Finally, the Mg K-edge in GX~5-1 suffers from a lower signal to noise ratio with respect to the Si K-edge, 
mainly because of the relatively high column density value.
However, the Mg K-edge can in principle help significantly in breaking model degeneracies.

%*************************************************************************************************************************

\subsection{Scattering and particle size distributions}
We use a MRN distribution to describe the particle size distribution. The distribution with particles sizes ranging between
$0.005\,\mu\mathrm{m}$ and $0.25\,\mu\mathrm{m}$ implies that most of the mass is in the large particles, while most of the surface area is 
in the small particles.  
MRN offers a simple parameterization, useful in this study.
It is beyond the scope of this paper to test more sophisticated models
such as \citet{Weingartner01} and \citet{Zubko04}. 
We note, however, that these size distributions have a maximum size cutoff around $0.25\,\mu\mathrm{m}$ 
and the silicate-type distributions do not differ dramatically from one another.

\begin{figure}
 \begin{center}
 \includegraphics[scale=0.5]{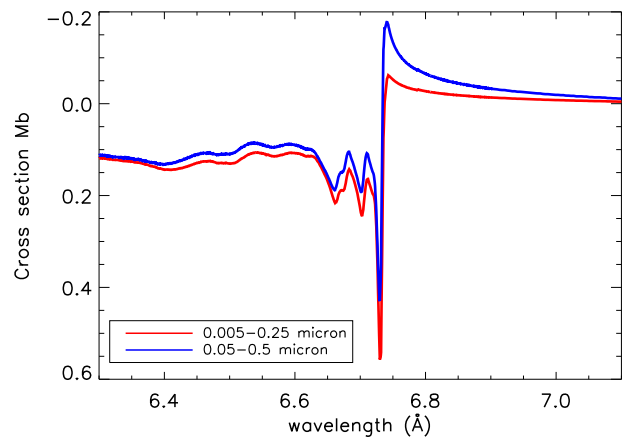}
 \caption{\small{The effect of a change in the size distribution on the extinction. The olivine Si K-edge with an MRN distribution 
 with particle sizes of $0.005-0.25\,\mu\mathrm{m}$ micron is shown in red. The blue line shows the same edge, but now the particle sizes range between 
 $0.05-0.5\,\mu\mathrm{m}$.}}
 \label{fig:largerpart}
  \end{center}
\end{figure}

In Figure~\ref{fig:fit}, it can be observed that there is a feature between $6.8\AA$ and the edge at $6.7\,\AA$. 
The MRN distribution predicts a large amount of small particles in our models. 
These particles do not add significantly to 
the scattering features of the extinction profiles.
The model does not fit the data in the area around $6.7\,\AA$, well, 
which may indicate the presence of larger particles along the line of sight than present in our model \citep{Hulst1958}.
This section therefore contains an investigation of the effect of the particle size distribution and focuses, in particular,
on the presence of particles larger than $0.25\,\mu\mathrm{m}$.  
In order to model the enhancement of the scattering peak around $6.7\AA$ we introduce a range of particle sizes: 
$0.05-0.5\,\mu\mathrm{m}$. 
The effect of a change in the size distribution is shown in Figure~\ref{fig:largerpart}. The olivine Si K-edge with a MRN distribution 
with particle sizes of $0.005-0.25\,\mu\mathrm{m}$ is shown in red and in blue the same edge is shown but now with a MRN size distribution that has
a particle range of $0.05-0.5\,\mu\mathrm{m}$. 
 
% Figure can be found in ~/idl/idl_prog/spex_prog/scattering_diff/make_plot_scat_olivine.pro

\begin{figure}
 \begin{center}
 \includegraphics[scale=0.5,left]{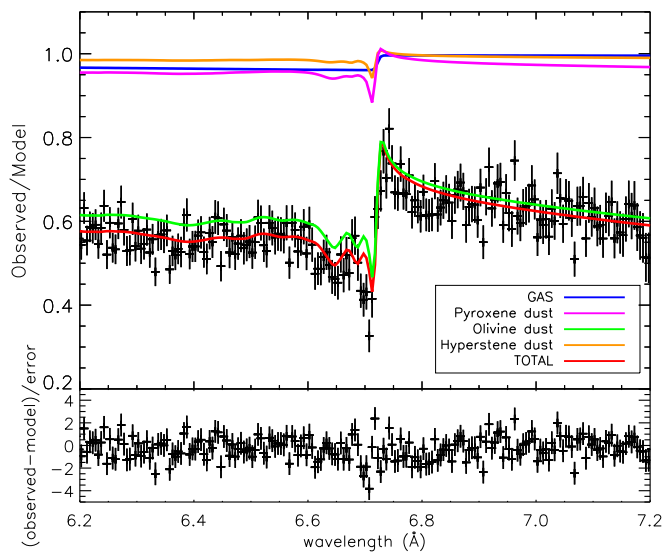}
%  \vspace{10pt}
 \caption{\small{The upper panel shows a fit of the Si K-edge with particles of sizes in a range of $0.1-0.5\,\mu\mathrm{m}$. The best fitting mixture
 consist out of the same compounds as the fit in section 6.2: sample 1 (crystalline olivine) and sample 5 (amorphous pyroxene) with a 
 small addition of sample 6 (hypersthene). The lower panel shows the models residuals of the fit in terms of the standard deviation, $\sigma$.}}
 \label{fig:largerpartfit}
  \end{center}
\end{figure}

The edge is fitted again with models of the same compounds, that contain the new particle distribution. 
We use the same parameters as in the first fit in section 6.2. The results are shown in Figure \ref{fig:largerpartfit}.
The reduced $C^2$ value of the best fit is: 1.08.  
There is a small contribution (of less than $2\%$ of the total amount of dust) of hypersthene (sample 6) in this 
fit in addition to amorphous pyroxene (sample 5) and crystalline olivine (sample 1). The contribution is so small that it does not significantly change
any of the results derived above. The silicon abundance remains at $4.0\pm0.4\times10^{-5}$ per H atom. 
The scattering feature before the edge is better fitted using a model with larger particles, which leads to the possibility
of the presence of particles larger than $0.25\,\mu\mathrm{m}$. 
It is possible that in environments such as the Galactic center region we observe a 
substantial amount of particles larger than $0.25\,\mu\mathrm{m}$ \citep{Ossenkopf92,Ormel09,Ormel11}. 
In our analysis, we assume that the dust particles are solid spheres,
while it is more likely that large particles in dense environments are grown by coagulation of dust particles~\citep{Jura80}. 
Porous dust particles show an enhanced extinction profile when compared to solid particles of the same mass due to their 
larger surface area. Solid particles are too massive to be abundant enough to cause the enhanced scattering.
Another indicator that dust particles are non-spherical is the observation of polarized starlight. Spheroidal
grain models are able to reproduce these observations \citep{Kim95,Draine09} and allow for larger particles in the size 
distribution \citep{Draine09}.
The effects of large porous grains in X-rays have also been studied by \citet{Hoffman16}. They conclude that 
grains with a significant porosity produce narrower forward scattering peaks than equal-mass non-porous grains.
\citet{Chiar06} analyzed the 
possibility of the presence of solid and porous spherical dust particles along sight lines toward four Wolf Rayet stars. 
They conclude
that a mixture of solid and porous silicates fits the $9.7\,\mu\mathrm{m}$ and $18\,\mu\mathrm{m}$ absorption features.
The presence of large porous dust particles along the line of sight toward GX~5-1 could be an explanation for the observed 
prominent scattering features in the Si K-edge.
The presence of larger grains is also derived from studies of the mid-infrared extinction law.
The extinction curve in the diffuse ISM is represented by $R_V=3.1$, while higher values of $R_V$ (i.e., 4-6) 
are observed for dense clouds, which may indicate the presence of larger grains \citep{Weingartner01}.
\citet{Xue16} calculate the intrinsic mid-infrared color excess from the stellar effective temperatures in order to determine the mid-infrared
extinction. They find that the extinction curve is consistent with the $R_V=5.5$ model curve 
and agrees well with the WD01 \citep{Weingartner01} interstellar dust model.  
The sight line toward GX~5-1 traverses the molecular ring and likely probes a mixture of diffuse and dense medium. The dense region may be associated with
the molecular ring, characterized by larger grains \citep{Ormel09,Ormel11}. 

%figure created in: ~/idl/idl_prog/spex_prog/all_over_again/all_with_newmodels/second_ref/diff_scat_with_res/plot_best_fit_diff_scat.pro

\section{Summary}

In this paper, we analyze the X-ray spectrum of the low-mass X-ray binary GX~5-1, where we focuse, in particular,
on the modeling of the Si K-edge. 
The Si K-edges of six silicate dust samples were measured at the Soleil synchrotron facility in Paris. 
Using these new measurements, we calculated the extinction profiles of these samples in order to make them suitable for the analysis
of the ISM toward GX~5-1. The extinction profiles of the Si K-edge were added to the AMOL model of the spectral fitting 
program SPEX. 
We obtained a best fit to the Chandra HETG data of GX~5-1 and arrive at the following results:

   \begin{enumerate}
      \item We established conservative lower limits on the abundances of Si, O, Mg, and Fe: $A_{\mathrm{Si}}/A_{\sun}>1.14$, 
      $A_{\mathrm{O}}/A_{\sun}>1.06$, $A_{\mathrm{Mg}}/A_{\sun}>1.6$,  $A_{\mathrm{Fe}}/A_{\sun}>0.79$.  
      Except for iron, all the lower limits on the abundances show abundances similar to, or above, protosolar values.
      We obtained upper limits on the depletion: $<0.87$ for silicon, $<0.23$ for oxygen, $<0.97$ for magnesium and $<0.76$ for iron.
      \item There may be indications for large dust particles along the line of sight due to enhanced scattering features in the Si K-edge.
      The scattering feature longward of the Si K-edge is better fitted using a model with larger particles, 
      which indicates the presence of particles larger than 0.25\,$\mu$m up to 0.5\,$\mu$m.
      \item The sharp absorption features observed in the Si K-edge suggest a 
      possibly significant amount of crystalline dust with respect to the total amount of dust. 
      However, more laboratory measurements are required to draw any conclusion on this subject.  
      \end{enumerate}

\begin{acknowledgements}
Dust studies at SRON and Leiden Observatory are supported through
the Spinoza Premie of the Dutch science agency, NWO.
We would like to thank J\"{o}rn Wilms and the anonymous referee for providing us with helpful comments and suggestions. 
This research made use of the Chandra Transmission
Grating Catalog and archive (http://tgcat.mit.edu). Furthermore, we made use of the FLUO correction code provided by 
Daniel Haskel. H.M. and S.Z. are grateful for
support of the Deutsche Forschungsgemeinschaft
under Mu 1164/7-2 and Mu 1164/8-1.
\end{acknowledgements}

\begin{appendix}
\section{Correction for saturation}
In section 3, we describe the effect of saturation. In order to correct for saturation, we used the FLUO correction code
by Daniel Haskel, which is part of the UWXAFS software package \citep{Stern1995117}. 
In this appendix, we give a brief summary of the correction method, which can also be found in the documentation of the FLUO code. 
Equation~\ref{eq:flou2} corresponds to pre-edge subtracted data, that is, for 
energies lower than the edge $I_f/I_0$, which is zero. The signal can be normalized by performing an edge-step normalization. The 
normalized signal $N$ is given by:
\begin{equation}
 N= \dfrac{\frac{I_f}{I_0}(E)}{\frac{I_f}{I_0}(E_0^{+})}=\Bigg[\dfrac{\epsilon_f(E)\alpha_e(E)}{\epsilon_f(E_0^{+})\alpha_e(E_0^+)}\Bigg]\Bigg[\dfrac{\alpha_{\mathrm{tot}}(E_f)+\alpha_b(E_0^+)+\alpha_e(E_0^+)}{\alpha_\mathrm{tot}(E_f)+\alpha_b(E)+\alpha_e(E)}\Bigg],                                                                  
\end{equation}

where $\alpha_\mathrm{tot}=\alpha_b+\alpha_e$, $\alpha_e$ is the absorption from the element of interest, $\alpha_b$
denotes the absorption from all other atoms and other edges of interest and $E_0^{+}$ indicates the energy above the main absorption edge. 
Dividing the denominator by $\alpha_e(E_0^{+})$ gives:

\begin{equation}
 N = \dfrac{\alpha_e(E)}{\alpha_e(E_0^{+})}\dfrac{\epsilon_f(E)}{\epsilon_f(E_0^{+})} \Bigg[\dfrac{\frac{\alpha_{\mathrm{tot}}(E_f)}{\alpha_e(E_0^{+})}+\frac{\alpha_b(E_0^{+})}{\alpha_e(E_0^{+})}+\frac{\alpha_e(E_0^+)}{\alpha_e(E_0^+)}}{\frac{\alpha_{\mathrm{tot}}(E_f)}{\alpha_e(E_0^+)}+\frac{\alpha_b(E)}{\alpha_e(E_0^+)}+\frac{\alpha_e(E)}{\alpha_e(E_0^{+})}}\Bigg].
\end{equation}

Defining $\beta=\frac{\alpha_{\mathrm{tot}}(E_f)}{\alpha_e(E_0^{+})}$, $\gamma=\frac{\alpha_b(E)}{\alpha_e(E_0^{+})}$, and $\gamma'=\frac{\alpha_b(E_0^{+})}{\alpha_e(E_0^{+})}$ and solving for $\frac{\alpha_e(E)}{\alpha_e{E_0^{+}}}$ gives:

\begin{equation}
 \dfrac{\alpha_e(E)}{\alpha_e(E_0^{+})} = \dfrac{N(\beta+\gamma)}{(\beta+\gamma'+1)-N}.
 \label{eq:appendix}
\end{equation}

Since we are only concerned with a small energy range at the interval of the XANES region, the following approximation was
made: $\frac{\epsilon_f(E)}{\epsilon_f(E_0^{+})}\approx1$ in Equation~\ref{eq:appendix}. Furthermore, 
$\alpha_b(E)\approx\alpha_b(E_0^{+})$ which leads to: $\gamma\approx\gamma'$. FLUO uses tabulated cross-section from the McMaster tables
to calculate $\beta$ and $\gamma$. It will also perform an edge-step normalization in order to obtain $N$. In this way
$\dfrac{\alpha_e(E)}{\alpha_e(E_0^{+})}$ can be derived.

\section{Si K-edge models}
Here we show the extinction profiles around the Si K-edge for the compounds used in this analysis.  
These profiles are implemented in the AMOL model of the fitting code SPEX.
The numbers of the samples correspond to the 
numbers given in Table~\ref{table:elements} in which the chemical formulas and the structure (crystalline or amorphous) are given.
The absolute cross-sections of the models used in this analysis are available in tabular form on the following website:
\url{www.sron.nl/~elisa/VIDI/}. 

\begin{figure}
 \begin{center}
 \includegraphics[scale=0.6, right]{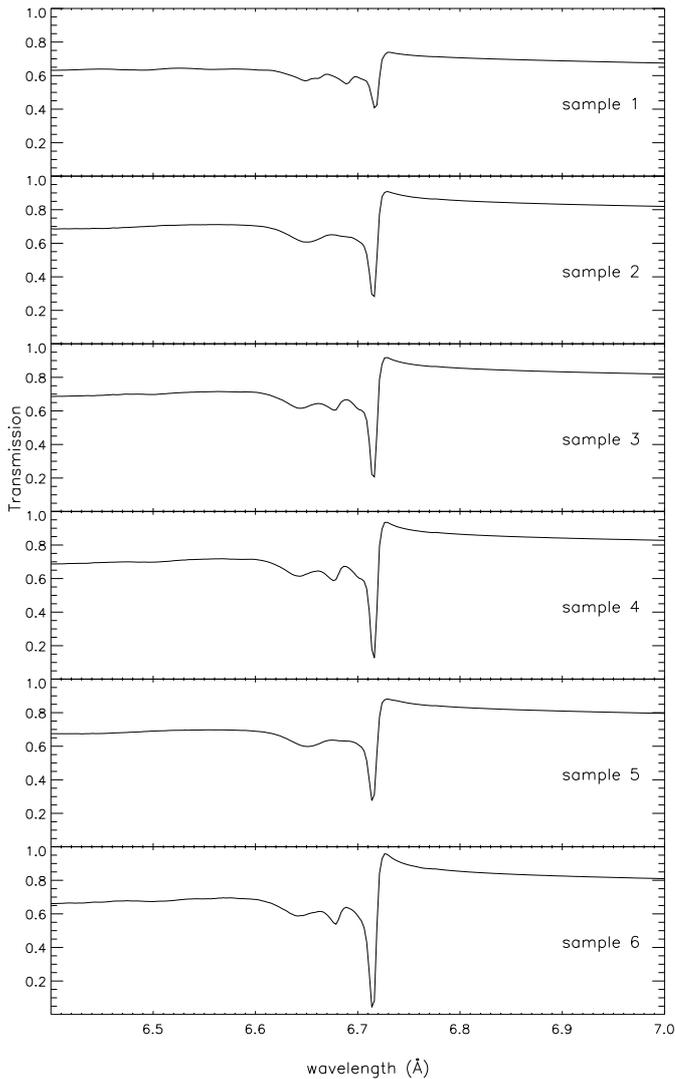}
 \vspace{0.5pt}
 \caption{\small{Transmission of the six dust extinction models (absorption and scattering) from the analysis of the Si K-edge. The silicon column density has been set here to 
 $10^{18}\,\mathrm{cm}^{-2}$ for all the dust models. Each model is indicated by a number which correspond to the numbers in
 Table~\ref{table:elements}. The table also gives the chemical formula and the structure (crystalline or amorphous) of the dust compound.}}
 \label{fig:all_models}
  \end{center}
\end{figure}
% Figure can be found in ~/idl/idl_prog/spex_prog/all_over_again/all_with_newmodels/example_models_plot/plot_models.pro

\end{appendix}

%-------------------------------------------------------------------

% \bibliography{zeegers}

\end{document}